\documentclass[pdftex,twocolumn,epjc3]{svjour3}

\RequirePackage[T1]{fontenc}
\smartqed  % flush right qed marks, e.g. at end of proof

\RequirePackage{graphicx}
\RequirePackage{mathptmx}
\RequirePackage{flushend}
\RequirePackage[numbers,sort&compress]{natbib}
\RequirePackage[colorlinks,citecolor=blue,urlcolor=blue,linkcolor=blue]{hyperref}

\journalname{Eur. Phys. J. C}

\graphicspath{{./PDF/}}

\usepackage[below]{placeins}
\usepackage{flafter}
\usepackage{amsmath}
\usepackage{mathtools}
\usepackage{enumitem}
\usepackage{overpic}
\usepackage{tabularx}
\usepackage{upgreek}
\usepackage{dblfloatfix}
\usepackage{microtype}
\usepackage{notoccite}
\usepackage{comment}
\DeclareMathAlphabet{\mathcal}{OMS}{cmsy}{m}{n}

\usepackage{tikz}
\usetikzlibrary{positioning}
\usetikzlibrary{calc}
\definecolor{mylightred}{RGB}{211,79,73}
\definecolor{mydarkred}{RGB}{199,44,38}
\definecolor{mylightgreen}{RGB}{78,153,67}
\definecolor{mydarkgreen}{RGB}{43,129,33}
\definecolor{mylightpurple}{RGB}{150,107,178}
\definecolor{mydarkpurple}{RGB}{126,78,160}
\definecolor{mylightblue}{RGB}{49,101,205}
\definecolor{mydarkblue}{RGB}{20,92,205}
\tikzset{juliadot/.style args={#1,#2}{shape=circle,line width=0.03ex,minimum width=0.4ex,fill=#1,draw=#2}}
\newcommand\julialetter[1]{{\strut\fontfamily{cmss}\bfseries\selectfont{#1}}}
\DeclareRobustCommand\julia{%
\begin{tikzpicture}[baseline=0mm, every node/.style={inner sep=0mm, outer sep=0mm}]
\node[anchor=base]        (j) at (0,0) {\julialetter{\j}};
\node[anchor=base, right=0ex of j] (u) {\julialetter{u}};
\node[anchor=base, right=0ex of u] (l) {\julialetter{l}};
\node[anchor=base, right=0ex of l] (i) {\julialetter{\i}};
\node[anchor=base, right=0ex of i] (a) {\julialetter{a}};
\path let \p1 = (j) in node[juliadot={mylightblue,mydarkblue}] (bluedot) at (\x1+0.02ex,1.4ex) {};
\path let \p1 = (i) in node[juliadot={mylightred,mydarkred}] (reddot) at (\x1,1.4ex) {};
\path let \p1 = (reddot) in node[juliadot={mylightpurple,mydarkpurple}] (purpledot) at (\x1+0.5ex,\y1) {};
\path let \p1 = (reddot) in node[juliadot={mylightgreen,mydarkgreen}] (greendot) at (\x1+0.25ex,\y1+0.42ex) {};
\end{tikzpicture}%
}

\newcommand{\SSD}{\mbox{\textit{SolidStateDetectors.jl}}}
\newcommand{\Cs}{661.660}
\newcommand{\Ba}{356.017}
\newcommand{\Baone}{276.398}
\newcommand{\Batwo}{302.852}
\newcommand{\Bathree}{356.017}
\newcommand{\Bafour}{383.851}

\usepackage[switch]{lineno}
\nolinenumbers

\begin{document}

\title{A novel wide-angle Compton Scanner setup to study bulk events in germanium detectors}

\author{I.~Abt\thanksref{a} \and C.~Gooch\thanksref{a} \and F.~Hagemann\thanksref{a,e} \and L.~Hauertmann\thanksref{a} \and D.~Hervas Aguilar\thanksref{a2,a3,a} \and X.~Liu\thanksref{a} \and O.~Schulz\thanksref{a} \and M.~Schuster\thanksref{a} \and A.J.~Zsigmond\thanksref{a}}

\thankstext{e}{e-mail: hagemann@mpp.mpg.de (corresponding author)}
\institute{Max-Planck-Institut für Physik, F\"ohringer Ring 6, 80805 Munich, Germany\label{a} \and Department of Physics and Astronomy, University of North Carolina, 120 E. Cameron Ave., Phillips Hall CB3255, Chapel Hill, 27599 NC, USA\label{a2} \and Triangle Universities Nuclear Laboratory, 116 Science Drive, Duke University, Durham, 27708 NC, USA\label{a3}}

\date{Received: date / Accepted: date}

\maketitle

\begin{abstract}
A novel Compton Scanner setup has been built, commissioned and operated at the Max-Planck-Institute for Physics in Munich to collect pulses from bulk events in high-purity germanium detectors for pulse shape studies. In this fully automated setup, the detector under test is irradiated from the top with \Cs\,keV gammas, some of which Compton scatter inside the detector. The interaction points in the detector can be reconstructed when the scattered gammas are detected with a pixelated camera placed at the side of the detector. The wide range of accepted Compton angles results in shorter measurement times in comparison to similar setups where only perpendicularly scattered gammas are selected by slit collimators. In this paper, the construction of the Compton Scanner, its alignment and the procedure to reconstruct interaction points in the germanium detector are described in detail. The creation of a first pulse shape library for an n-type segmented point-contact germanium detector is described. The spatial reconstruction along the beam axis is validated by a comparison to measured surface pulses. A first comparison of Compton Scanner pulses to simulated pulses is presented to demonstrate the power of the Compton Scanner to test simulation inputs and models.
\end{abstract}

%--------------------------------------------------------

\section{Introduction} \label{sec:introduction}

High-purity germanium detectors find applications in a variety of fundamental research fields. They are used to measure nuclear structures~\citep{AGATA2012,GRETINA2013} and coherent elastic neutrino-nucleus scattering~\citep{vGeN2015,CONUS2019}, as well as to search for dark matter~\citep{CoGeNT2013,SuperCDMS2014,CDEX2018} and neutrinoless double-beta decay~\citep{GERDA2020,MAJORANA2019,LEGEND2017,LEGEND2021}. In these experiments, the time development of the charges induced on the contacts of the detectors, i.e.\;the pulses, are recorded for all events above a given energy threshold. Analyzing the pulse shapes provides information about the event topologies, which allows to reconstruct the positions of charge carrier generation~\citep{AGATA2021} or to identify background in low-background~experiments~\citep{GERDA2013,MAJORANA2019_PSA}.

Algorithms for pulse shape analysis can be tuned by testing their efficiencies on a selection of pulses with known origin. These so-called pulse shape libraries can either be simulated using a dedicated software package~\citep{SigGen,Bruyneel2016,Abt2021}, or created from data by depositing energy in well-defined volumes in the detector. In the latter case, the germanium detector is often irradiated with a collimated beam consisting of alphas, betas, or low-energy gammas~\citep{GALATEA,TUBE,CAGE}, which deposit their energy within less than a few millimeters away from the surface. This allows for the investigation of surface effects and dead layers but does not provide information on pulses from the inner parts of the~detector.

One way of creating a pulse shape library for the bulk of the detector is based on the Compton effect~\citep{Petry1993,Vetter2000,Abt2008,Dimmock2009,Ha2013,vonSturm2017}. A common approach~\citep{Vetter2000,Dimmock2009,Ha2013,vonSturm2017} is to irradiate the detector under test with gammas from a collimated $^{137}$Cs source. Perpendicularly scattered gammas which pass through horizontal slit collimators are registered by energy-sensitive detectors placed nearby. The pulses for different volumes of the detector are recorded, step by step, by moving the collimated source and the slit collimators to different positions. This approach has been very successful in detector bulk characterization campaigns. A drawback, however, is the restriction to perpendicularly scattered gammas; this either limits the statistics of the measurement or results in very long measurement~times.
Another technique of collecting pulses from bulk events in highly segmented detectors~\citep{Crespi2008,DeCanditiis2020} is based on specific pulse shape comparison procedures when irradiating the detector with collimated gamma beams from different~sides.

In this paper, a novel and compact Compton Scanner~setup equipped with an energy- and position-sensitive camera is~presented. This setup does not require any collimation of the scattered gammas. The increased range of usable Compton angles results in higher statistics and shorter measurement~times.

%--------------------------------------------------------

\section{Construction} \label{sec:construction}

\subsection{Basic working principle} \label{subsec:workingprinciple}

\begin{figure}[b]
    \centering
    \includegraphics[width=\linewidth]{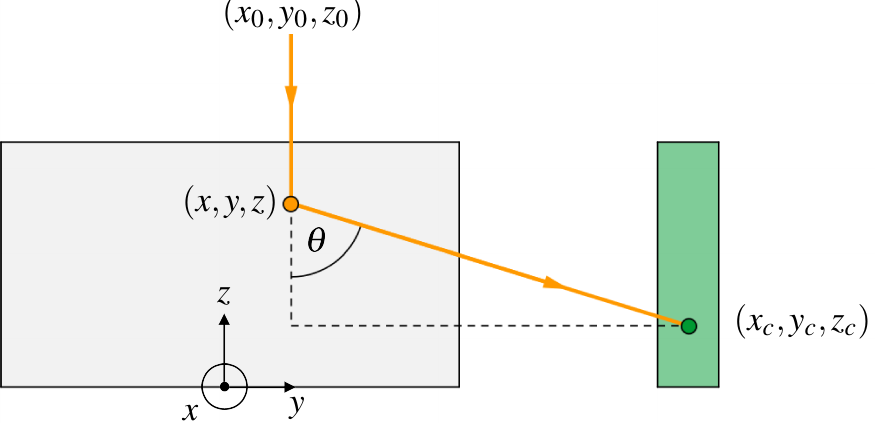}
    \caption{Schematic of a typical gamma trajectory in the Compton Scanner: The gamma is emitted by the source at $(x_0, y_0, z_0)$, Compton scatters inside the germanium detector at $(x,y,z)$ and is fully absorbed in the camera at $(x_c, y_c, z_c)$. The Cartesian coordinate system is defined such that the $y$~axis is parallel to the line which connects the centers of the detector and the camera, and the $z$~axis is parallel to the beam. The origin of the Cartesian coordinate system is located at the center of the bottom surface of the detector.}
    \label{fig:workingprinciple}
\end{figure}

In the Compton Scanner, the detector under test is irradiated vertically from the top with a collimated gamma beam. A schematic of a typical gamma trajectory is shown in Fig.~\ref{fig:workingprinciple}. The figure also depicts the Cartesian coordinate system, $x$, $y$ and $z$, used throughout this paper. Gammas which deposit part of their energy, $E_\text{det}$, through Compton scattering in the detector are deflected by the Compton angle, $\theta$, with 
\begin{linenomath}
\begin{equation} \label{eq:compton}
\cos(\theta) = 1 - \dfrac{m_ec^2 E_\text{det}}{E_\text{in} \left(E_\text{in} - E_\text{det}\right)}~,
\end{equation}
\end{linenomath}
where $E_\text{in}$ is the energy of the incoming gammas.

If a scattered gamma is fully absorbed in the camera placed nearby, which measures both energy and position, the location of the scattering point in the detector, $(x,y,z)$, can be reconstructed. Assuming there is exactly one scattering point on axis, i.e.\;$x = x_0$ and $y = y_0$, where $(x_0,y_0,z_0)$ is the position of the center of source, the unknown $z$~coordinate can be calculated as
\begin{linenomath}
\begin{equation} 
z = z_\theta \coloneqq z_c + \sqrt{(x_c-x_0)^2 + (y_c-y_0)^2} \;\cot(\theta)~, \label{eq:ztheta}
\end{equation}
\end{linenomath}
where $z_\theta$ is the $z$ coordinate of the scattering point as reconstructed from the Compton angle $\theta$ and from the position of the hit in the camera, $(x_c, y_c, z_c)$. The relation $z = z_\theta$ in Eq.\;\eqref{eq:ztheta} is true as long as the beam is parallel to the $z$ axis.

Pulse shape libraries are created by moving the source to different positions $x_0$ and $y_0$. Note that with this setup, for any position of the source, data are collected for the whole range in $z$ at the same time.

\subsection{Requirements} \label{subsec:requirements}

The purpose of this Compton Scanner setup is to create pulse shape libraries from bulk events with an intrinsic spatial resolution of less than $\pm1\,\text{mm}$ in all dimensions. This volume is compatible with the volume of the charge clouds at energies around 2\,MeV as expected in events from neutrinoless double-beta decay~\citep{Abt2007}. This results in a set of physical requirements on the components of the Compton Scanner:
\begin{itemize}[noitemsep,topsep=0.5\baselineskip]
    \item The gamma source should be monoenergetic, with a single gamma line at an energy that has a high penetration power into germanium and that has a high probability for Compton scattering. The gammas should predominantly scatter into an interval around 90\textdegree.
    \item The diameter of the collimated beam needs to be less than 2\,mm throughout the detector volume to reach the target resolution in the $x$ and $y$ directions. The collimation of gammas with energies suitable for Compton scatter requires several centimeters of high-density material~\citep{Thoraeus1965}, e.g.\;lead or tungsten.
    \item For each position of the source, the data taking should only take up to a few hours, such that multiple detectors per year can be scanned at multiple temperatures. For this, a highly active source with an activity of several hundred MBq is required.
    \item  The camera needs to have a spatial and energy resolution good enough to reconstruct $z$ using Eq.\;\eqref{eq:ztheta} with an uncertainty of less than $\pm1\,\text{mm}$.
    \item The Compton Scanner setup is to accommodate arbitrary detector cryostats, demanding it to be modular and to have a framework to mount and move the components. Enough space in the center of the setup is required where the detector cryostat can be installed.
\end{itemize}
In addition, the reconstruction based on Eq.\;\eqref{eq:ztheta} requires that the positions of the three main components of the Compton Scanner, i.e.\;source, detector and camera, are all known and aligned within a common coordinate system.

\subsection{Components} \label{subsec:components}

\begin{figure*}[tbh]
    \centering
    \begin{overpic}[width=\linewidth]{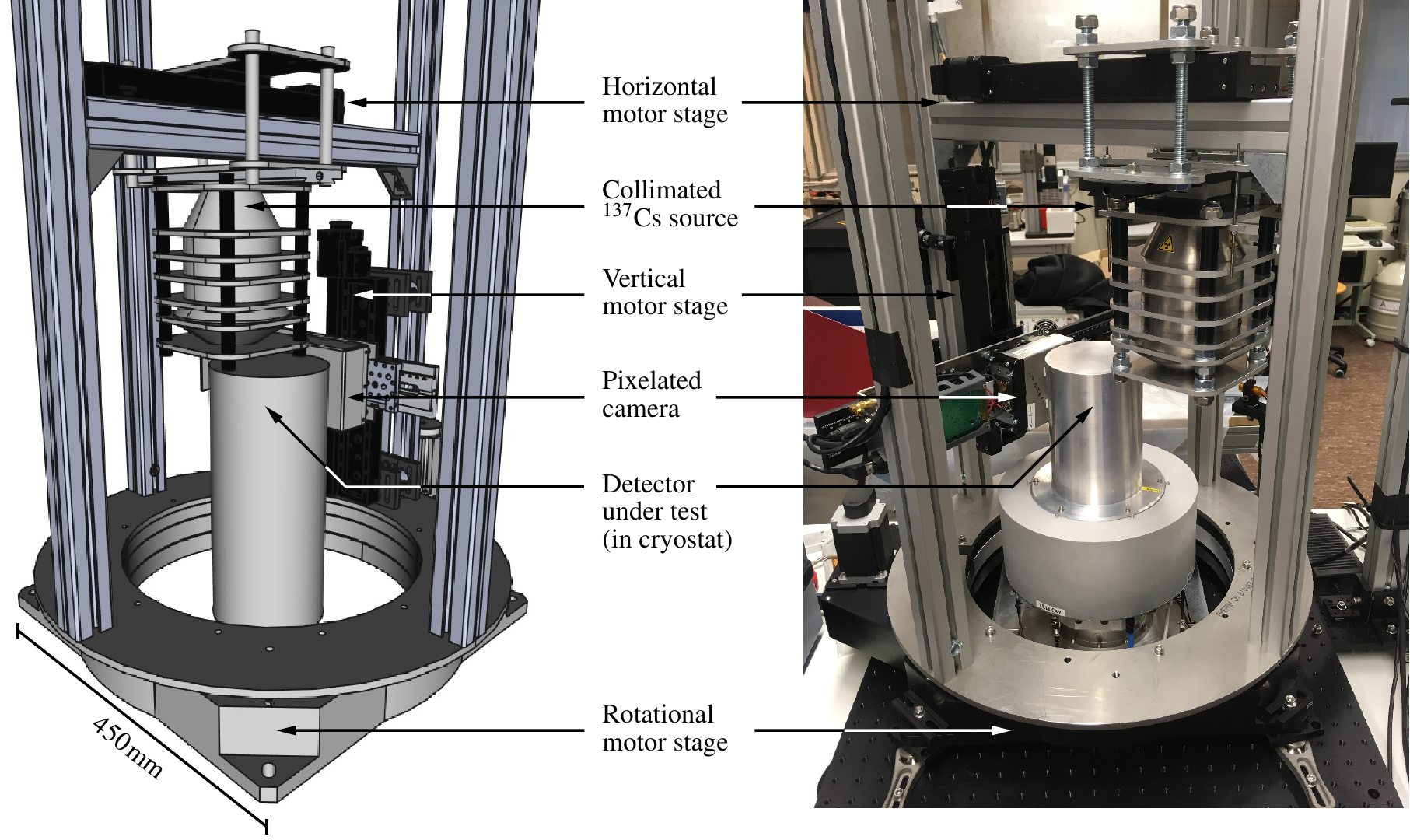}
    \bfseries
    \put(19,-1){\makebox(0,0){(a)}}
    \put(78.5,-1){\makebox(0,0){(b)}}
    \end{overpic}\\[10pt]
    \caption{(a)~3D model and (b)~photo of the Compton Scanner setup. The vertical motor is used to place the pixelated camera at an optimal height with respect to the detector under test. The horizontal motor allows to move the $^{137}$Cs source along the line which connects the centers of the detector cryostat and the pixelated camera. The rotational motor is used to rotate the whole frame, along with the pixelated camera and the source, around the cryostat.}
    \label{fig:comptonscanner}
\end{figure*}

The individual components are mounted on a common frame as shown in Fig.~\ref{fig:comptonscanner}. 
The frame consists of a circular base plate with four vertical MayTec aluminum profiles, two of which are connected via a fifth horizontal profile. The source and the camera are attached to a horizontal and a vertical translation motor stage (STANDA\textsuperscript{\tiny\textregistered} 8MT50-200BS1-MEn1), respectively, which allow to move them with a precision of 5\,µm. The base plate is fixed onto a rotational motor stage (STANDA\textsuperscript{\tiny\textregistered} 8MRB450-360-60-MEn2), which allows to rotate it with a precision of 0.15\textdegree. The rotational motor stage is mounted on a $600\,\text{mm} \times 600\,\text{mm}$ aluminum breadboard, %(THORLABS\textsuperscript{\tiny\textregistered} MB6060/M),
with a $300\,\text{mm} \times 300\,\text{mm}$ cutout in the center, which is located on a table with a circular hole. The detector in its cryostat is inserted through this hole. The Compton Scanner can accommodate cryostats with diameters of up to 130\,mm. Two line lasers, attached to two adjacent vertical MayTec profiles, are used to roughly align the center of the Compton Scanner frame with respect to the center of the detector~cryostat.

The source used in the Compton Scanner is a cylindrical 740\,MBq $^{137}$Cs source with an active diameter of 0.9\,mm. It emits gammas with a characteristic energy of \Cs\,keV~\citep{Firestone} and fulfills the requirements listed in Sect.\ \ref{subsec:requirements}. The emission of the gammas is isotropic. To create a narrow vertical beam, the source is embedded in a collimator.

\begin{figure}[tbp]
    \centering
    \includegraphics[width=0.93\linewidth]{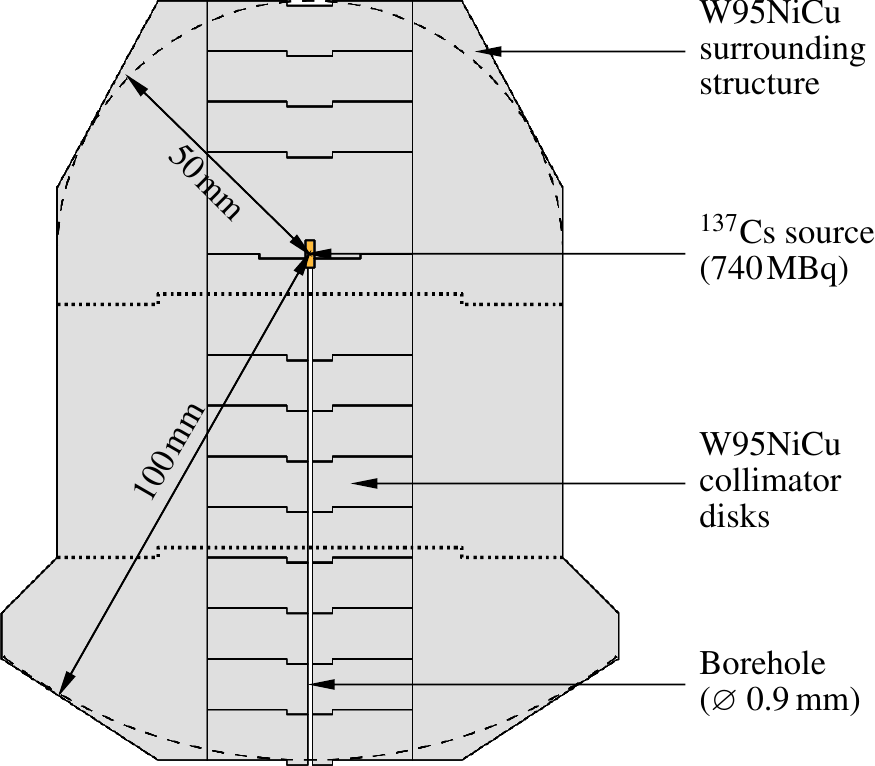}
    \caption{Schematic cross section of the rotationally symmetric $^{137}\text{Cs}$ source collimator.}
    \label{fig:sourcecollimator}
\end{figure}

The collimator used in the Compton Scanner is shown in Fig.~\ref{fig:sourcecollimator}. 
\begin{figure*}[!b]
    \vspace{-10pt}
    \centering
    \begin{overpic}[width = \textwidth]{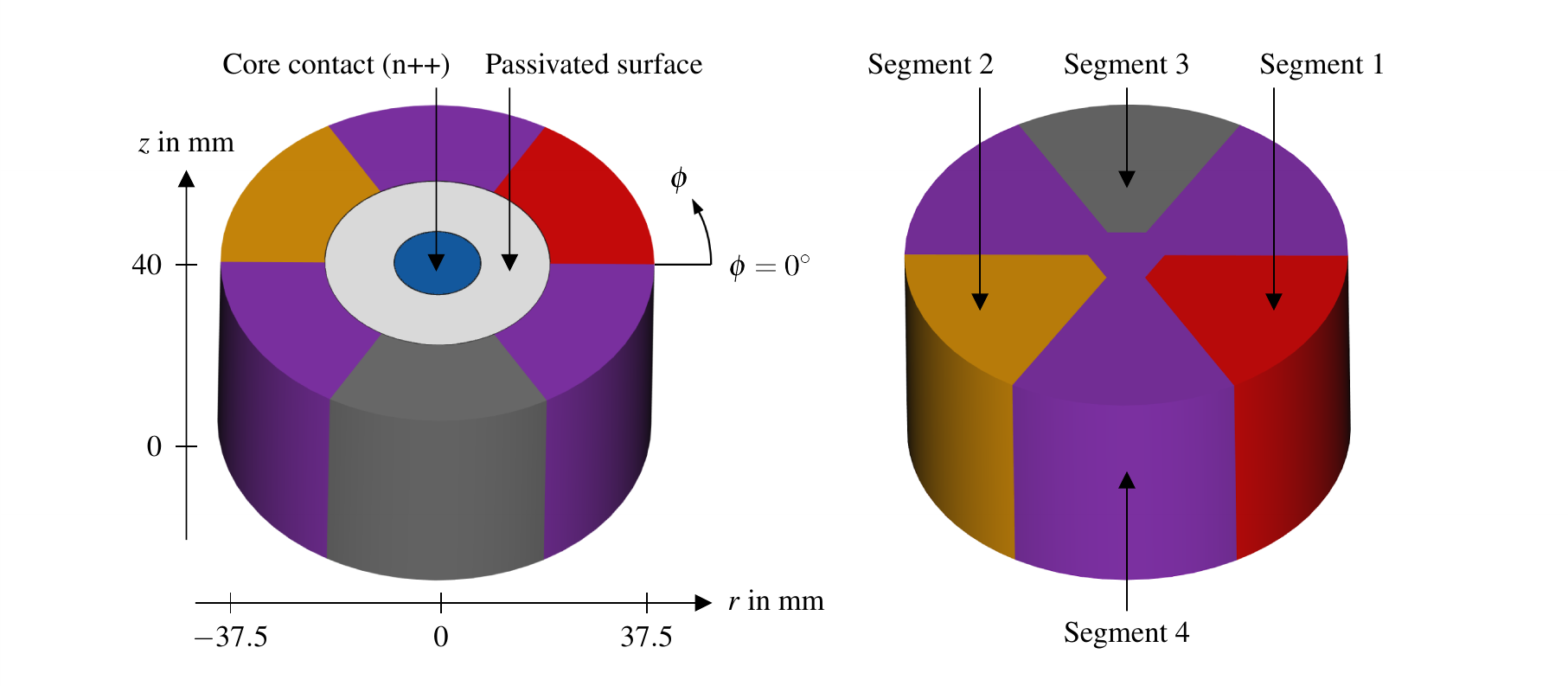}
    \bfseries
    \put(28.125,0){\makebox(0,0){(a)}}
    \put(71.875,0){\makebox(0,0){(b)}}
    \end{overpic}
    \nocite{NIST,H3D,Zhu2011,He1996}
    \caption{Schematic of the n-type segBEGe detector~\citep{Abt2019} used in the first Compton Scanner measurement campaign: (a) top view and (b) bottom view. In (a), the cylindrical coordinate system to describe positions in the segBEGe detector is shown.}
    \label{fig:segBEGe}
\end{figure*}
It consists of 13 disks and a three-piece surrounding structure made out of W95NiCu tungsten alloy. Nine of the disks form a 100\,mm long, massive collimator with a borehole of 0.9\,mm in diameter.  At the top, the collimator assembly provides a shielding of at least 50\,mm of tungsten for radiation safety. The geometry of the collimator ensures that downwards but not vertically emitted 661.660\,keV gammas have to pass through at least 100\,mm of tungsten. This reduces the fraction of downward gammas that do not pass through the borehole to less than~$10^{-7}$. The diameter of the borehole leads to an aperture angle of the collimated beam of~0.26\textdegree. The collimated source is mounted as close to the detector cryostat as possible in order to minimize the width of the beam reaching the detector. If the detector inside the cryostat is placed within 100\,mm underneath the bottom of the collimator, the beam spot diameter is less than~1.8\,mm.

The mean free path of \Cs\,keV gammas in germanium is around 26.9\,mm~\cite{NIST}. For perpendicularly scattered gammas, 373\,keV are deposited in the germanium detector. The scattered 289\,keV gammas have a mean free path of 16.8\,mm~\citep{NIST}. The charge collection times in germanium detectors are typically between 400\,ns and 1000\,ns, depending on the geometry and operational parameters.

The camera used to detect the Compton scattered gammas is a customized OEM module produced by H3D, Inc.~\citep{H3D}. The camera system consists of four pixelated cadmium-zinc-telluride (CdZnTe) detectors. The mean free path of 289\,keV gammas in CdZnTe is around 11.2\,mm~\cite{NIST}. Each detector is based on a $22\times22\times10\,\text{mm}^3$ CdZnTe crystal. The four detectors are mounted inside an aluminum box in a $2\times2$ configuration with~2\,mm gaps between them. For each CdZnTe detector, the $22\times22\,\text{mm}^2$ surface facing the germanium detector is fully covered by the cathode. The back $22\times22\,\text{mm}^2$ surface is covered by the anode which consists of $11\times11$ pixels with a center-to-center distance of~1.9\,mm. This contact geometry allows for a three-dimensional reconstruction of the positions of the energy depositions (hits) in the camera. The lateral and depth position resolution of the camera are both less than 0.5\,mm full width at half maximum, FWHM. The sub-pixel lateral resolution is reached by considering the effect of charge sharing between neighboring pixels~\citep{Zhu2011}. The depth resolution is achieved by comparing the anode and cathode pulse amplitudes~\cite{He1996}.

The camera readout system, trigger and reconstruction algorithms are proprietary technology of H3D, Inc.~\citep{H3D}. The camera was bought together with software providing a data stream consisting of deposited energy, time and position values for each event. The energy resolutions of the pixelated camera at energies relevant for Compton Scanner measurements were determined by irradiating the camera directly with gammas emitted from a $^{133}$Ba source, selecting only events with a single energy deposition. The resolution was observed to be homogeneous across the pixels, varying by less than 10\%. The average values for the measured FWHM of the characteristic gamma peaks of $^{133}$Ba are listed in~\mbox{Table~\ref{tab:cameraenergyresolution}.}

\begin{table}[tbph]
    \vspace{-12pt}
    \centering
    \caption{Absolute and relative energy resolutions of the pixelated camera used in the Compton Scanner, determined from events with the characteristic $^{133}$Ba gamma energies~\citep{Firestone} and only one energy deposition in the camera. The width of the peak, $\Delta E$, corresponds to the~FWHM.}\label{tab:cameraenergyresolution}
    \begin{tabular}{c|cc}
        $E$ in keV & $\Delta E$ in keV & $\Delta E/E$ \\ \hline
        \Baone   & 3.02 & 1.09\% \\
        \Batwo   & 3.09 & 1.02\% \\
        \Bathree & 3.16 & 0.89\% \\
        \Bafour  & 3.26 & 0.85\%
    \end{tabular}
    \\[-36pt]
\end{table}

%--------------------------------------------------------

\section{Integration of a detector} \label{sec:integration}

\subsection{n-type segBEGe detector} \label{subsec:segBEGe}

The first detector operated in the Compton Scanner was an n-type segmented Broad Energy Germanium (segBEGe) detector~\citep{Abt2019}. A schematic of this detector is depicted in Fig.~\ref{fig:segBEGe}. The detector has a diameter of 75\,mm and a height of 40\,mm. It has one n++ point contact, the so-called core contact, and a four-fold-segmented p++ contact geometry on the mantle. This segmentation results in five pulses per event, providing additional information about the event topology compared to an unsegmented detector. The core contact has a diameter of 15\,mm. It is surrounded by a ring covered with a passivation layer with an outer diameter of 39\,mm. The segments~1, 2 and~3 each extend over an arc of 60\textdegree{} and are evenly distributed every 120\textdegree. Segment~4 covers the remaining regions on the top and mantle surface and is closed at the \mbox{bottom end-plate}.

Positions in the segBEGe detector are described in a cylindrical coordinate system using the coordinates, $r$, $\phi$ and $z$, as shown in Fig.~\ref{fig:segBEGe}a. The center of the detector in the horizontal plane is defined as $r = 0\,\text{mm}$. The polar angle $\phi = 0^\circ$ is located at the boundary between the segments~1 and~4 which is closer to segment~3. Looking from the top, $\phi$ increases counterclockwise. The $z$~axis points from the bottom to the top of the detector, with $z = 0\,\text{mm}$ located at the bottom of the detector.

For measurements with the Compton Scanner, the n-type segBEGe detector was installed in the temperature-controlled cryostat K2~\citep[p.\,39]{Hagemann2020}. It was operated at the recommended bias voltage of 4500\,V using an iseg SHQ226L high voltage power supply~\citep{isegSHQ226L}. The charges induced on the core and all segment contacts were amplified using charge-sensitive preamplifiers (EURISYS PSC823C~\citep{PSC823}) and recorded using a 14-bit analog-to-digital converter (STRUCK SIS3316-250-14~\citep{STRUCK}) ADC, with a sampling frequency of 250\,MHz, i.e.\;a sampling time of 4\,ns. All pulses from the core contact, which collects electrons, were inverted by the ADC in order to record them with positive pulse amplitudes.

\subsection{Intrinsic spatial resolution} \label{subsec:intrinsicresolution}

To accommodate the K2 cryostat, the pixelated camera was placed such that its surface had a minimal distance of 72\,mm to the center of the Compton Scanner frame. The vertical centers of the segBEGe detector and the camera were aligned. The latter maximizes the number of detectable perpendicularly scattered gammas. According to the Klein-Nishina distribution~\cite{Klein1929}, placing the camera at lower $z$ would increase the overall number of detectable gammas but shift the selection of gammas to smaller Compton angles, resulting in higher uncertainties on the $z$ reconstruction. The distance between the top of the detector and the bottom of the source collimator was 31\,mm.

\begin{figure*}[tb]
    \centering
    \begin{overpic}[width=\linewidth]{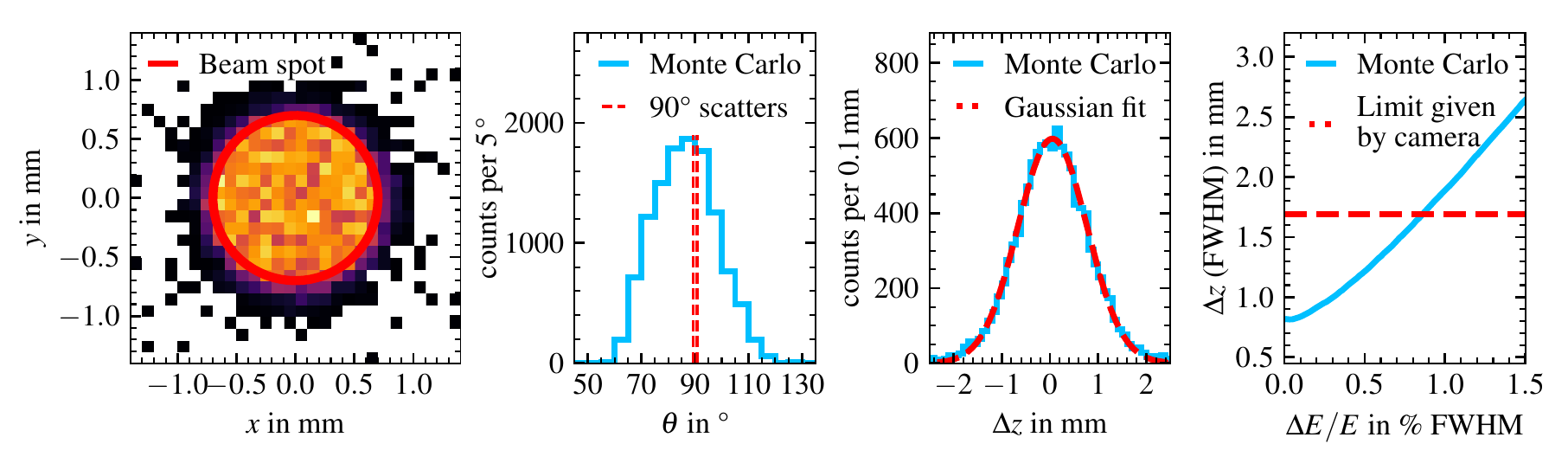}
    \put(0,30){\makebox(0,0)[l]{$r = 0\,\text{mm}$:}}
    \end{overpic}
    \begin{overpic}[width=\linewidth]{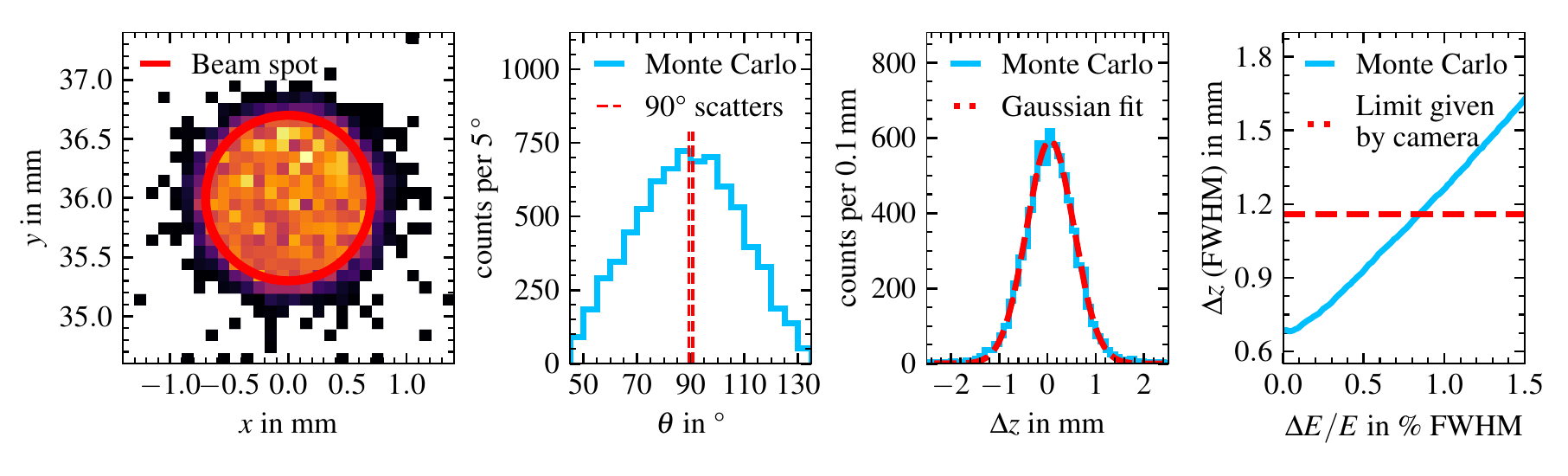}
    \put(0,30){\makebox(0,0)[l]{$r = 36\,\text{mm}$:}}
    \bfseries
    \put(18.5,0.5){\makebox(0,0){(a)}}
    \put(44.1,0.5){\makebox(0,0){(b)}}
    \put(66.6,0.5){\makebox(0,0){(c)}}
    \put(89.5,0.5){\makebox(0,0){(d)}}
    \end{overpic}
    \nocite{Hagemann2020, isegSHQ226L, PSC823, STRUCK, Klein1929}
    \caption{(a)~Beam spot profile as determined from a Monte Carlo simulation performed with the simulation toolkit \textsc{Geant4}~\citep{GEANT4}. (b)~Distribution of Compton angles, $\theta$, of Monte Carlo events with only one Compton scatter in the germanium detector. Shown as a dashed band is the angle range of perpendicularly scattered gammas, \mbox{$\theta = (90.00\pm0.72)^\circ$}, expected for a 80\,mm long and 1\,mm wide collimator slit and perfect collimation. (c)~Deviation, $\Delta z$, of the reconstructed $z_\theta$ using Eq.\;\eqref{eq:ztheta} from the true $z$ in the simulation using the energy and spatial resolutions from the camera as listed in Sect.~\ref{subsec:intrinsicresolution}. (d)~FWHM of the $\Delta z$ distribution, $\Delta z$ (FWHM), as a function of the energy resolution, $\Delta E/E$, of the detector at \Ba\,keV. The dashed line in~(d) depicts the upper limit for the precision of the reconstruction, achieved when the energy reconstructed by the camera is used. The minimal distance from the surface of the pixelated camera to the vertical beam is 72\,mm (top) and 36\,mm (bottom), respectively.}
    \label{fig:intrinsic_alignment}
\end{figure*}

The intrinsic spatial resolution for this configuration of the Compton Scanner was determined from Monte Carlo simulations performed with the simulation toolkit \textsc{Geant4}~\citep{GEANT4}. Two scenarios with different positions of the vertical beam were simulated:
\begin{enumerate}[noitemsep, label=(\roman*)]
    \item At $r = 0\,\text{mm}$, i.e.\;the vertical beam placed at the center of the detector and the Compton Scanner frame.
    \item At $r = 36\,\text{mm}$, i.e.\;the vertical beam placed 36\,mm away from the center of the Compton Scanner frame at a minimal distance of 36\,mm to the surface of the camera.
\end{enumerate}
Only events with exactly one Compton scatter in the germanium detector and at least one hit in the pixelated camera were selected. The lateral and depth position resolutions of the camera of 0.5\,mm FWHM were used in the Monte Carlo simulation for smearing the hit positions. The energy resolution of the camera of 0.89\% FWHM at \Ba\,keV was added as uncertainty on the hit energies. The $z$~coordinates were reconstructed for all selected events using Eqs.\;\eqref{eq:compton}~and~\eqref{eq:ztheta}. The results are shown in Fig.~\ref{fig:intrinsic_alignment}.

The size of the beam spot in the $xy$-plane does not depend on the $xy$-position of the source. The beam spot diameter at the top (bottom) of the detector was 1.23\,mm (1.61\,mm) for all radii, resulting in an average of 1.40\,mm, see Fig.~\ref{fig:intrinsic_alignment}a. This agrees well with the diameters expected from the aperture of the beam.

Figure~\ref{fig:intrinsic_alignment}b depicts the distribution of Compton angles for all simulated events with a single scatter in the germanium detector. At \mbox{$r = 36\,\text{mm}$}, the gamma beam is closer to the camera, resulting in the solid angle of the acceptance being four times larger than for \mbox{$r = 0\,\text{mm}$}. Thus, a significant number of events with $\theta$ below 60\textdegree{} and above 120\textdegree{} are~detected.

During measurements, the Compton angles cannot be measured directly but are calculated from the energies deposited in the detector or in the camera. For cases in which a germanium detector has a poor energy resolution, $E_\text{det}$ can be determined from the energy deposited in the camera, $E_\text{cam}$, via \mbox{$E_\text{det} = E_\text{in} - E_\text{cam}$} with a resolution given by the uncertainties on $E_\text{cam}$. The resulting Gaussian distributions of the differences, $\Delta z$, between the reconstructed $z_\theta$ and the true $z$ for this case are shown in Fig.~\ref{fig:intrinsic_alignment}c. The resolution for reconstructing $z$ is 1.69\,mm FWHM at $r = 0\,\text{mm}$ and 1.16\,mm FWHM at $r = 36\,\text{mm}$. This defines an upper limit for the intrinsic $\Delta z$ resolution of the Compton Scanner setup.

Alternatively, $E_\text{det}$ can be determined directly from the core pulses recorded from the germanium detector. If the energy resolution of the detector is better than the resolution on $E_\text{det}$ determined via $E_\text{cam}$, the intrinsic spatial resolution is further improved. The dependence of the $\Delta z$ resolution on the energy resolution of the detector is shown in Fig.~\ref{fig:intrinsic_alignment}d. For the n-type segBEGe detector, the energy resolution at \Bathree\,keV is 0.74\% FWHM~\citep{Abt2019}. When reconstructing $\theta$ using the energy deposited in the segBEGe detector, the $\Delta z$ resolution is slightly improved to 1.52\,mm FWHM at $r = 0\,\text{mm}$ and to 1.09\,mm FWHM at $r = 36\,\text{mm}$.

If the same $\Delta z$ resolution were to be targeted with a similar setup using a single 80\,mm long slit collimator, the width of the collimator slit would need to be below~1\,mm. Assuming perfect collimation, only gammas with Compton angles of \mbox{$(90.00\pm0.72)^\circ$} would pass through the collimator. This would reduce the fraction of usable gammas to \mbox{$(2.73\pm0.22)\%$} at $r = 36\,\text{mm}$ and to \mbox{$(4.30\pm0.19)\%$} at $r = 0\,\text{mm}$, see Fig.~\ref{fig:intrinsic_alignment}b. In addition, pulses would only be collected for 1\,mm out of the 40\,mm in $z$ for a given collimator position. This would further reduce the fraction of detectable gammas to \mbox{$(0.068\pm0.005)\%$} at $r = 36\,\text{mm}$ and to \mbox{$(0.107\pm0.005)\%$} at $r = 0\,\text{mm}$, resulting in a decrease in statistics of around 1000.
%$1471\pm103$ at $ r = 36\,\text{mm}$ and of $935\pm44$ at $r = 0\,\text{mm}$. 
Even with a higher acceptance in solid angle and multiple slits at different heights, the decrease in statistics compared to the Compton Scanner presented in this paper would be of the order of 100.

\subsection{Spatial alignment} \label{sec:alignment}

The first step to approach the intrinsic spatial resolution of the Compton Scanner was a mechanical alignment of the centers of the detector and of the Compton Scanner frame. After inserting the K2 cryostat through the hole in the table, the center alignment was of the order of a few millimeters. Line lasers, which produced a laser cross at the center of the Compton Scanner frame, were used to further align the centers. Moving the rotational motor stage on the aluminum breadboard until the laser cross matched the center of the cryostat allowed for a center alignment to several hundreds of micrometers.

\begin{figure}[!tbp]
    \centering
    \vspace{0.5cm}
    \includegraphics[width=\linewidth]{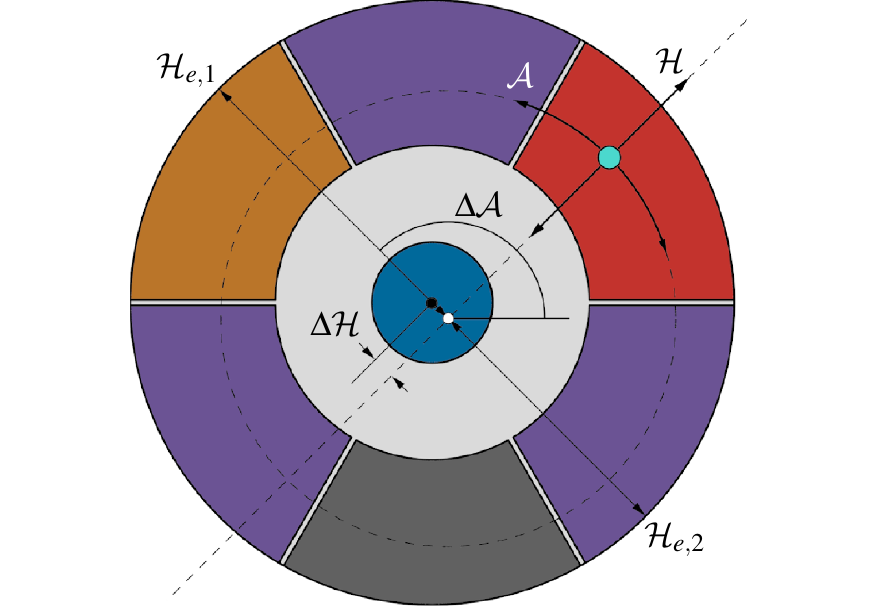}\\[20pt]
    \caption{Schematic of the movement of the source in the local motor coordinate system of the Compton Scanner, $\mathcal{H}$ and $\mathcal{A}$. Also shown are the misalignment quantities, $\Delta\mathcal{H}$ and $\Delta\mathcal{A}$, that define the misalignment between the center of the Compton Scanner frame (white point) and the center of the segBEGe detector (black point).}
    \label{fig:misalignment}
\end{figure}

The position of the source was set by the horizontal stage position, $\mathcal{H}$, and the rotational stage angle, $\mathcal{A}$, in the local motor coordinate system shown in Fig.~\ref{fig:misalignment}. The origin of this cylindrical coordinate system was located at the center of the Compton Scanner frame. The position of the source had to be expressed in the polar coordinates, $r$ and $\phi$, used in the segBEGe detector coordinate system introduced in Fig.~\ref{fig:segBEGe}a. If the centers were perfectly aligned, the source position, $\mathcal{H}$, in the local motor coordinate system and $r$ in the detector coordinate system would only differ by an offset, i.e.\;\mbox{$r = \mathcal{H} - \mathcal{H}_0$}. The same would apply to the rotational stage angle, $\mathcal{A}$, and the polar angle $\phi$, i.e.\;\mbox{$\phi = \mathcal{A} - \mathcal{A}_0$}. However, the unavoidable mismatch of the centers made the conversion from local motor coordinates to detector coordinates non-trivial and required further alignment measurements.

Figure~\ref{fig:misalignment} introduces the quantities that define a possible misalignment of the center of the Compton Scanner frame with respect to the center of the segBEGe detector. For a non-zero mismatch, $\Delta\mathcal{H}$, the detector edge is reached at different horizontal stage positions, $\mathcal{H}_e$, for different $\mathcal{A}$. There exists a rotational stage angle, $\Delta\mathcal{A}$, at which $\mathcal{H}_e$ reaches its maximum value,~$\mathcal{H}_{e,1}$. Accordingly, the minimum, $\mathcal{H}_{e,2}$, is reached after a rotation of 180\textdegree. The misalignment quantities, $\Delta \mathcal{H}~=~\frac12 (\mathcal{H}_{e,1}~-~\mathcal{H}_{e,2})$ and $\Delta\mathcal{A}$, as well as the offset, $\mathcal{H}_0~=~\frac12(\mathcal{H}_{e,1}~+~\mathcal{H}_{e,2})~+~R$, where $R$ is the radius of the segBEGe detector, were determined from data.

An alignment scan was performed to determine $\Delta\mathcal{H}$ and $\Delta\mathcal{A}$. In this scan, $\mathcal{A}$ was increased from 0\textdegree{} to~360\textdegree{} in steps of~5\textdegree{}. For every $\mathcal{A}$, the horizontal motor stage was moved over a range of 11\,mm in steps of 1\,mm across the edge of the detector. At each point, a 1\,min measurement was taken.

\begin{figure}[tb]
    \centering
    \includegraphics[width=\linewidth]{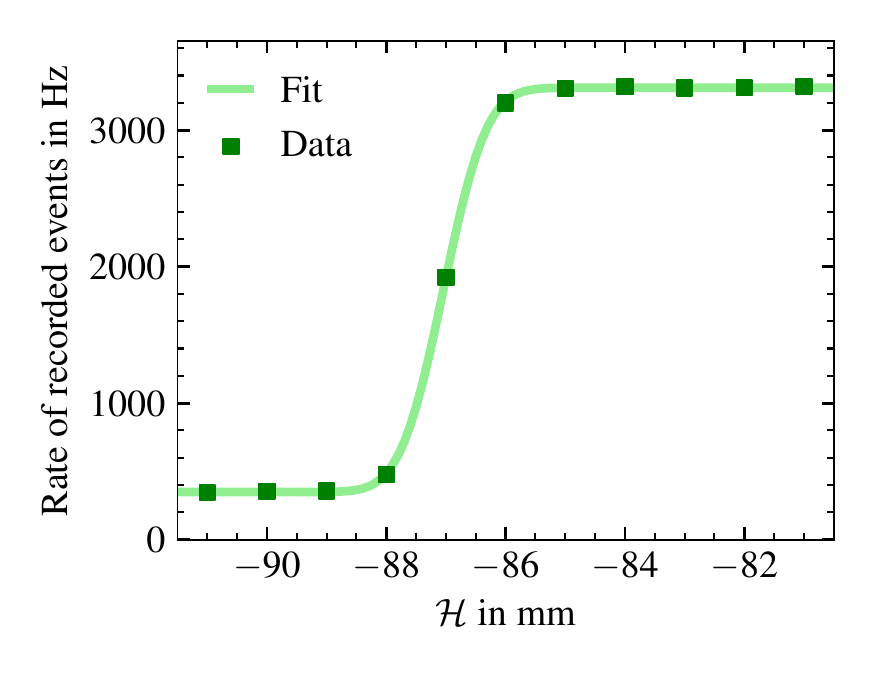} \\[-10pt]
    \caption{Rate of events in the segBEGe detector as determined from 1\,min measurements as a function of the horizontal motor position, $\mathcal{H}$, at $\mathcal{A} = 0^\circ$. The statistical uncertainties would be hidden by the markers and are not~shown.}
    \label{fig:hedetermination}
\end{figure}

Figure~\ref{fig:hedetermination} shows the rate of events, $N$, recorded in the segBEGe detector as a function of $\mathcal{H}$ for $\mathcal{A} = 0^\circ$. At low $\mathcal{H}$, the vertical beam does not intersect with the detector and only background events are observed. At high $\mathcal{H}$, the beam is fully contained within the detector volume, resulting in a higher count rate. The detector edge can be determined by fitting the data to a Gauss error function, i.e.
\begin{linenomath}
\begin{equation}
    N(\mathcal{H}) = \dfrac{N_0}{2} + \dfrac{N_0}{2} \text{erf}\left(\dfrac{\sqrt{2} (\mathcal{H} - \mathcal{H}_e(\mathcal{A}))}{R}\right) + B,
\end{equation}
\end{linenomath}
where $N_0$ is the event rate expected when the beam spot is fully contained on the detector surface and $B$ is the rate of background events. The rate of events from the $^{137}$Cs~source was measured to be around 3\,kHz, while the rate of background events was around 350\,Hz.

\begin{figure}[tb]
    \centering
    \includegraphics[width=\linewidth]{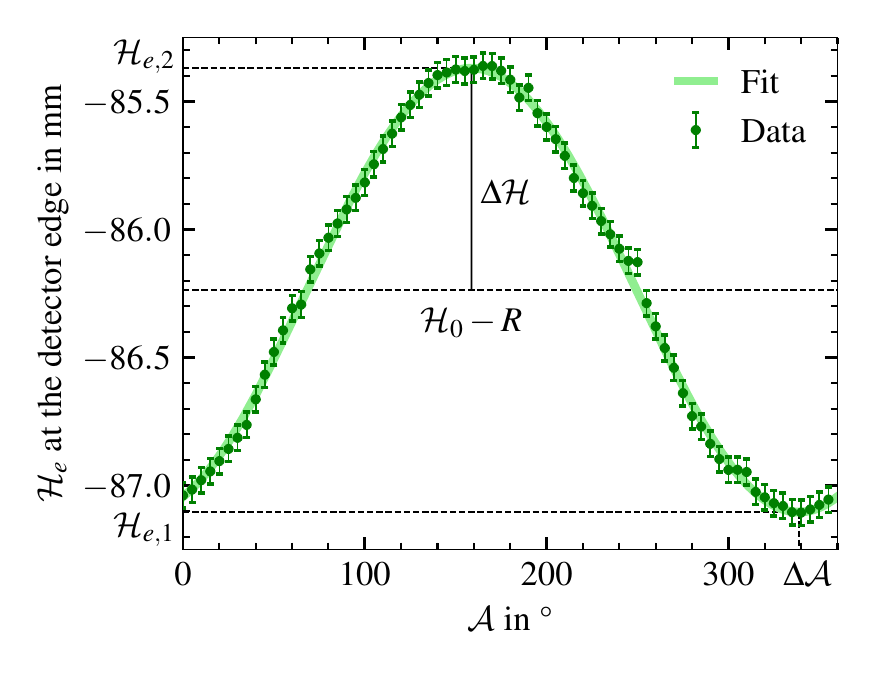} \\[-10pt]
    \caption{Horizontal stage position at which the detector edge is determined, $\mathcal{H}_e$, as a function of the rotational stage angle~$\mathcal{A}$. The misalignment quantities $\Delta\mathcal{H}$ and $\Delta\mathcal{A}$ are indicated. The error bars correspond to the uncertainties obtained from the fits to determine $\mathcal{H}_e$.}
    \label{fig:topalignment}
\end{figure}

For cylindrical detectors, the relation between $\mathcal{H}_e$ and $\mathcal{A}$ as derived from the geometry depicted in Fig.~\ref{fig:misalignment}~is
\begin{linenomath}
\begin{equation}
\begin{split}
\mathcal{H}_e(\mathcal{A}) = \mathcal{H}_0 &- \Delta \mathcal{H} \cos(\mathcal{A} - \Delta\mathcal{A}) \\ &\pm \sqrt{R^2 - (\Delta \mathcal{H})^2 \sin^2(\mathcal{A} - \Delta\mathcal{A})}~. \label{eq:HeR}
\end{split}
\end{equation}
\end{linenomath}
Details on the derivation of Eq.\;\eqref{eq:HeR} are provided in~\mbox{\ref{app:alignment}.}

The results for the segBEGe detector, obtained from fitting the function in Eq.\;\eqref{eq:HeR} to the measured $\mathcal{H}_e$, are shown in Fig.~\ref{fig:topalignment}. The center of the Compton Scanner frame was reached at \mbox{$\mathcal{H}_0=(-48.7\pm0.6)\,\text{mm}$}. The mismatch of the centers was \mbox{$\Delta\mathcal{H}=(0.87\pm0.06)\,\text{mm}$} at \mbox{$\Delta\mathcal{A}=(338.7\pm0.9)^\circ$}. The conversion of local motor coordinates with $\mathcal{H} < \mathcal{H}_0$ to $r$ and $\phi$ results from geometrical calculations:
\begin{linenomath}
\begin{align}
    r &= \sqrt{\vert\mathcal{H} - \mathcal{H}_0\vert^2 + (\Delta\mathcal{H})^2 - 2 \Delta\mathcal{H} \vert\mathcal{H} - \mathcal{H}_0\vert \cos(\mathcal{A}-\Delta\mathcal{A})}, \label{eq:realr} \\ 
    \phi &= \text{asin} \left(\dfrac{\vert\mathcal{H} - \mathcal{H}_0\vert\sin(\mathcal{A} - \Delta\mathcal{A})}{r} \right) + \Delta\mathcal{A} - \mathcal{A}_0~, \label{eq:realphi}
\end{align}
\end{linenomath}
where $\mathcal{A}_0 = (158.1\pm1.2)^\circ$ is the offset between $\phi = 0^\circ$ and the zero of the rotational motor stage. The value of $\mathcal{A}_0$ was determined by measuring the position of the segment boundaries in rotational motor coordinates~\citep{Abt2019}. In the following, all motor stage positions are converted to detector coordinates using Eqs.~\eqref{eq:realr} and~\eqref{eq:realphi}.

%--------------------------------------------------------

\section{Data acquisition and analysis} \label{sec:datataking}

\subsection{Data taking}

\begin{figure}[t]
    \centering
    \includegraphics[width=\linewidth]{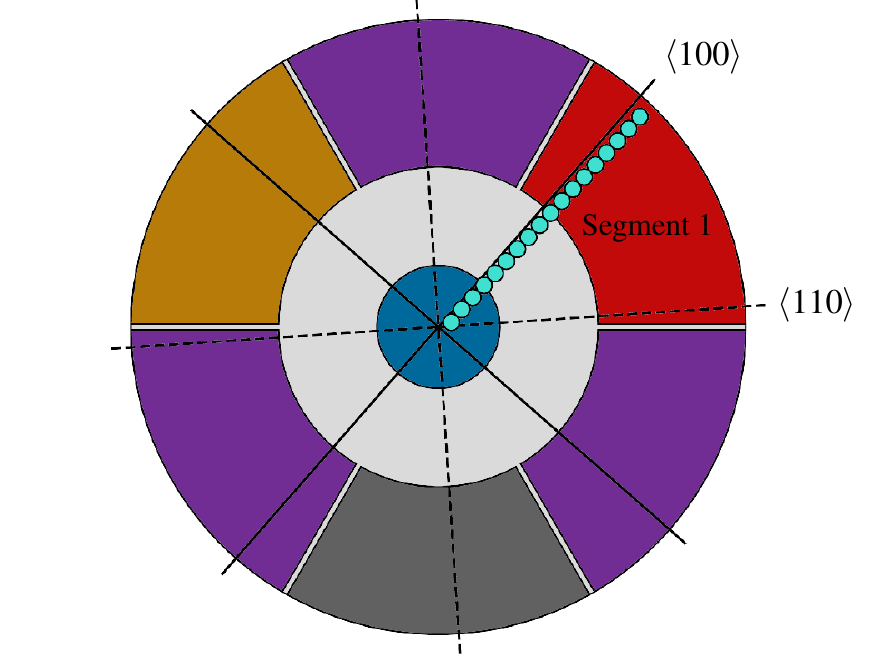}
    \setlength{\tabcolsep}{2pt}
    \begin{tabularx}{\linewidth}{@{}l|*{10}{>{\centering\arraybackslash}X}@{}}
    $r$ in mm  & 1.5   & 3.5  & 5.5  & 7.5  & 9.5  & 11.5 & 13.5 & 15.5 & 17.5 \\ \hline
    $t$ in min & 60  & 30 & 30 & 30 & 30 & 20 & 20 & 20 & 20 \\
    \multicolumn{3}{c}{}\\[-6pt]
    $r$ in mm  & 19.5 & 21.5 & 23.5 & 25.5 & 27.5 & 29.5 & 31.5 & 33.5 & 35.5\\ \hline
    $t$ in min & 20 & 20 & 10 & 10 & 10 & 10 & 10 & 10 & 10 \\
    \end{tabularx}
    \caption{Top view of the scan points and the measurement times, $t$, at the different radii,~$r$, used for the commissioning of the Compton Scanner. The solid and dashed black lines indicate the $\langle100\rangle$ and $\langle110\rangle$ crystallographic axes, respectively.}
    \label{fig:scanplan}
\end{figure}

The first set of measurements was performed by irradiating the detector along the $\langle100\rangle$ crystallographic axis in segment~1, see Fig.~\ref{fig:scanplan}. The scan points had a distance of 2\,mm. The measurement times were increased from 10 to 60\,min towards the center of the detector. This was done to mitigate the loss in statistics due to the higher probability of additional interactions of the scattered gammas in the germanium detector along their path to the camera. At all positions, the event rate in the segBEGe detector was around 3\,kHz. In the pixelated camera, the event rate was between 100\,Hz and~180\,Hz. 

The data acquisition system was not configured to build events at runtime. The two data streams from the germanium detector and from the pixelated camera were recorded independently and stored separately. For Compton Scanner measurements, only events with coincident energy depositions in the segBEGe detector and the pixelated camera are of interest. Both data streams were synchronized using 50\,ns long rectangular voltage pulses which were generated by an integrated circuit in the camera in time intervals of 2\,s and recorded by the ADC of the detector~\citep[p.\,44f]{Hagemann2020}.

After synchronization, the time stamps of coincident events in the detector and the camera typically coincided within a few microseconds. Detector and camera events with no respective counterpart within 50\,µs in the camera or detector were discarded. This resulted in a rate of coincident events of around 35\,Hz when irradiated in the center, and of around 140\,Hz when irradiated close to the edge of the~detector.

\subsection{Event selection}

\begin{figure}[tb]
    \centering
    \vspace{-12pt}
    \includegraphics[width=\linewidth]{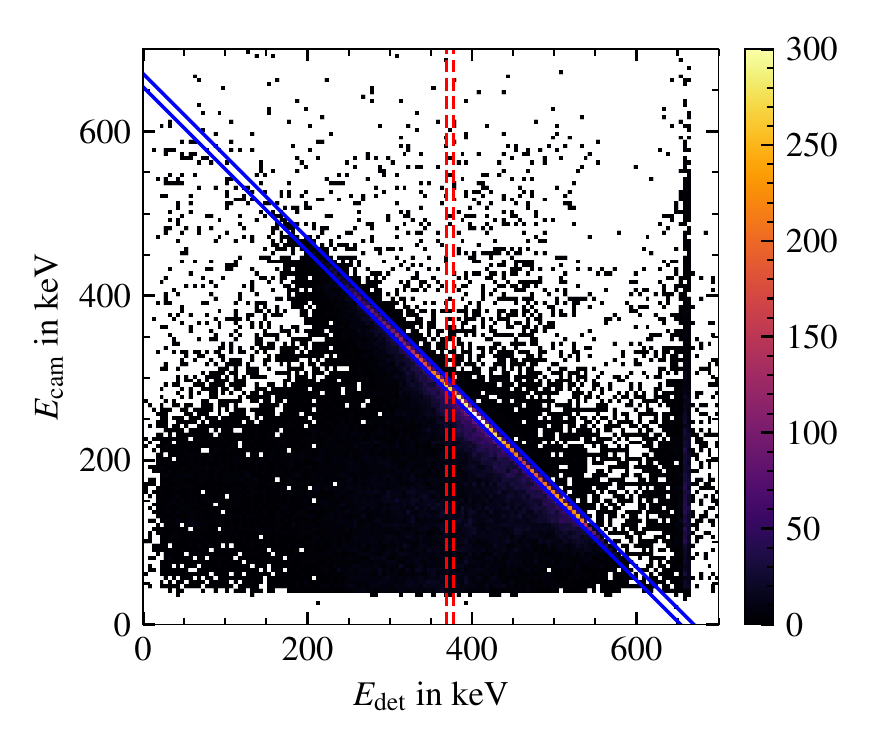}\\[-6pt]
    \caption{Distribution of energies measured in the core of the segBEGe detector, $E_\text{det}$, and in the pixelated camera, $E_\text{cam}$, for coincident events during a 10\,min Compton Scanner measurement on the $\langle100\rangle$ axis in segment~1 at \mbox{$r = 33.5\,\text{mm}$}. Both axes are divided into bins of 5\,keV. The solid lines indicate the region containing events with a sum energy of \mbox{$E_\text{det} + E_\text{cam} = (662\pm8)\,\text{keV}$.} The dashed lines indicate the energy range \mbox{$E_\text{det} = (373\pm4)\,\text{keV}$}, corresponding to perpendicularly scattered gammas in a setup with 80\,mm long and 2\,mm wide collimator slits.}
    \label{fig:energycut}
\end{figure}

\begin{figure*}[tb]
    \centering
    \begin{overpic}[width=\linewidth]{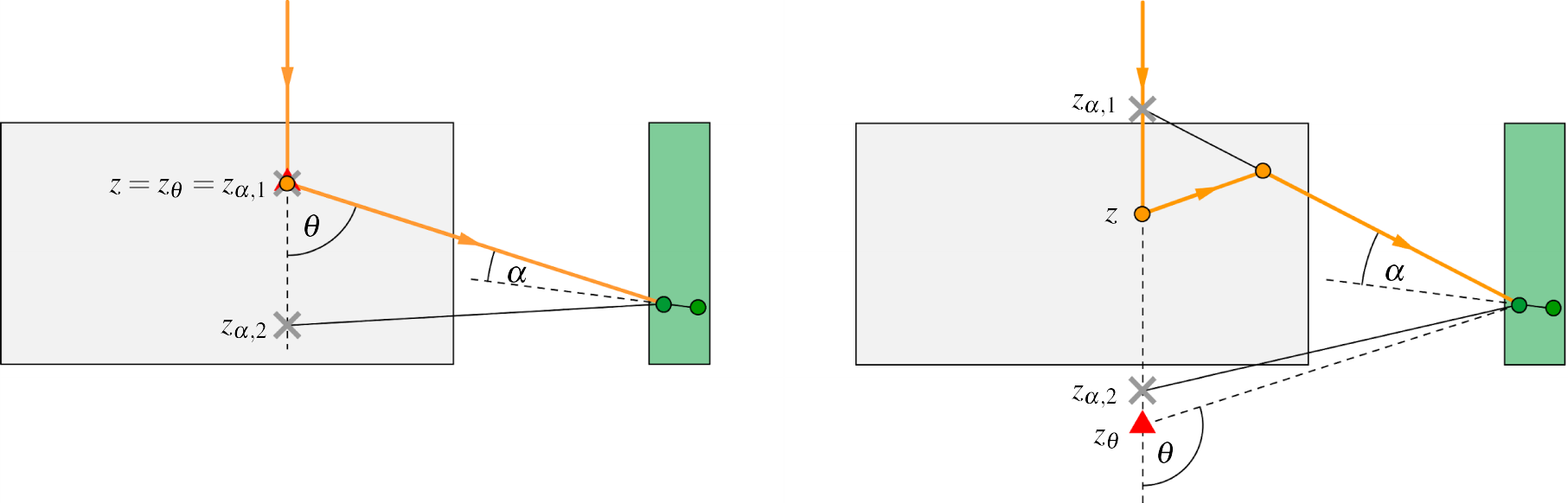}
    \bfseries
    \put(18.2,-2.5){\makebox(0,0){(a)}}
    \put(72.7,-2.5){\makebox(0,0){(b)}}
    \end{overpic}\\[20pt]
    \caption{Schematic of the trajectory of a fully-contained gamma for (a)~a single-site event and (b)~a multi-site event in the germanium detector. Interaction points in the germanium detector and in the camera are shown as dots, the reconstructed interaction point using Eq.\;\eqref{eq:ztheta} is shown as a triangle, and possible candidates obtained from the Compton cone using only camera information are shown as crosses. In (a), the dot, the triangle and one of the crosses overlap.}
    \label{fig:reconstruction}
\end{figure*}

The event selection for the creation of pulse shape libraries was performed using only coincident events. Events were selected if the energy in the segBEGe detector, $E_\text{det}$, as determined from the core pulse~\citep{Abt2019} and the total energy detected in the pixelated camera, $E_\text{cam}$, added up to the characteristic energy of the gammas from the $^{137}$Cs source, i.e.\;\mbox{$E_\text{det} + E_\text{cam} = (662 \pm 8)\,\text{keV}$}. This selection of fully-con\-tained events ensured that most of the events originated from gammas emitted by the $^{137}$Cs source and that the gammas did not undergo additional Compton scattering on their way from the segBEGe detector to the camera, e.g.\;off the cryostat wall or off the aluminum box of the pixelated camera.

Figure~\ref{fig:energycut} shows a correlation plot for $E_\text{det}$ and $E_\text{cam}$ obtained from a 10\,min measurement, for which the segBEGe detector was irradiated at $r = 33.5\,\text{mm}$, i.e.\;at a distance of 4\,mm to the edge. The diagonal line corresponds to fully-contained gammas from the $^{137}$Cs source and is densely populated. The fraction of fully-contained events varies from 10\% when irradiated in the center to 30\% when irradiated close to the detector edge. The lower fraction in the center arises from the higher probability that the scattered gamma is absorbed in the germanium detector. Figure~\ref{fig:energycut} also demonstrates the power of accepting a wide range of Compton angles. Restricting the selection to only events with perpendicularly scattered gammas, i.e.\;\mbox{$E_\text{det} = (373\pm4)\,\text{keV}$}, would further reduce the statistics by a factor of around 40.

\subsection{Data processing}  \label{subsec:reconstruction}

The reconstruction of $z$ presented in Sect.~\ref{subsec:workingprinciple} applies only to events in which the gamma Compton scatters exactly once in the detector, i.e.\;single-site events. Eq.\;\eqref{eq:ztheta} does not hold for events in which the gamma scatters multiple times in the detector, i.e.\;multi-site events. In these cases, the interaction point is misreconstructed and the pulses are attributed to the wrong~$z$.

Misreconstructed pulses can be identified for events with two well-separated hits in the pixelated camera. This approach was inspired by the procedures employed for Compton cameras used to image gamma sources~\citep{delSordo2009,Wahl2015,Kim2019}. The reconstruction of the origin of a gamma requires an assumption on which of the camera hits was the first hit, i.e.\;the Compton scatter. According to the Monte Carlo simulation, the first hit is expected to be the one with the lower energy. The Compton angle, $\alpha$, between the path of the gamma entering the camera and the line connecting both hits, see Fig.~\ref{fig:reconstruction}a, is calculated using Eq.\;\eqref{eq:compton}. This requires replacing $\theta$ by $\alpha$, $E_\text{in}$ by $E_{\text{cam},1} + E_{\text{cam},2}$, and $E_\text{out}$ by $E_{\text{cam},2}$, where $E_{\text{cam},1}$ and $E_{\text{cam},2}$ are the energy deposited in the first and the second hit in the camera, respectively. Geometrically, $\alpha$ and the line connecting the two hits define a Compton cone. The gamma entering the camera propagated along the surface of this Compton cone. Thus, the last interaction point of the gamma in the germanium detector has to lie on the surface of the Compton cone which intersects with the beam axis in up to two points, $z_{\alpha,1}$ and~$z_{\alpha,2}$.

For single-site events, see Fig.~\ref{fig:reconstruction}a, there is only one interaction point in the germanium detector and it is located on the beam axis. The surface of the Compton cone intersects with the beam axis in this interaction point, i.e.\;$z = z_\theta = z_\alpha$, where $z_\alpha$ is either $z_{\alpha,1}$ or $z_{\alpha,2}$. For multi-site events, see Fig.~\ref{fig:reconstruction}b, the last interaction point in the germanium detector is typically not located on the beam axis. The surface of the Compton cone passes through the last interaction point of the gamma in the germanium detector and intersects with the beam axis in so-called ghost points. These ghost points are often outside of the germanium detector and typically none of them agrees with the misreconstructed interaction point, i.e.\;$z_\theta \neq z_\alpha$.

Reconstructing $z_\theta$ as described in Sect.~\ref{subsec:workingprinciple} and demanding one of the $z_{\alpha,i}$ to agree with $z_\theta$ within 2\,mm allows to reject multi-site events without the need of pulse shape analysis. In the case that the reconstruction is rejected, the time order of the two hits is swapped. Only if both scenarios are rejected, the event is classified as a multi-site event and is discarded. The uncertainty of this validation is too large when the two hits in the camera are too close to each other. The method works reliably for events with a minimum distance of 3\,mm between the two hits. This distance is also larger than 1.5 times the center-to-center distance of the anode pixels which ensures that the camera can reliably separate multiple hits from each other.

%--------------------------------------------------------

\section{Creating a pulse shape library} \label{sec:pulseshapelibrary}

Pulses from individual events are too noisy to create a pulse shape library suitable to perform pulse shape analysis. Thus, in the order of 100 pulses for each detector volume are desirable to form so-called superpulses, i.e.\;averaged pulses, which are less affected by statistical fluctuations and electronic noise.

\begin{figure}[tp]
	\centering
	\includegraphics[width=\linewidth]{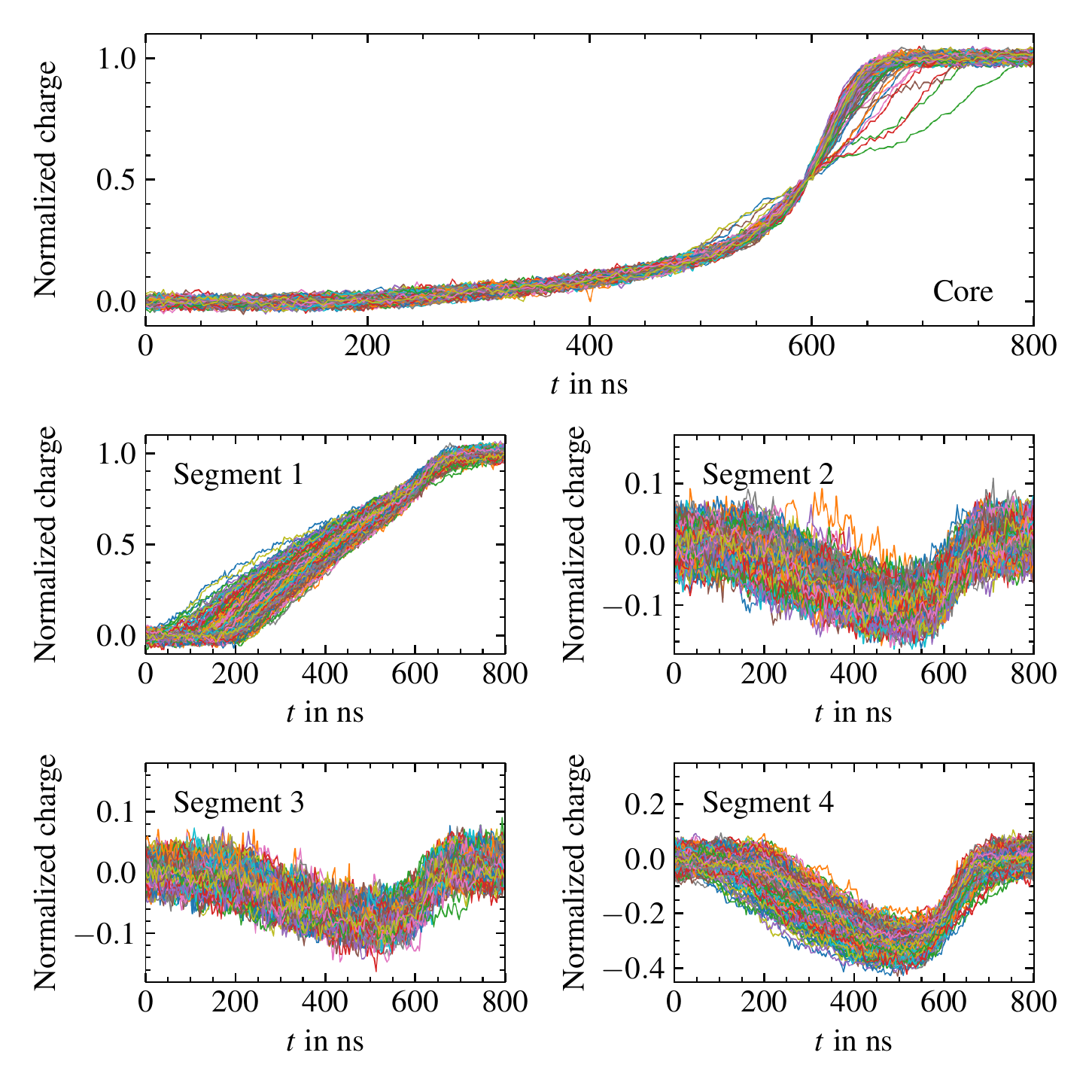}
	\caption{Individual pulses from events assigned to the interaction point at $r = 33.5\,\text{mm}$, $\phi = 46.1^\circ$ and a reconstructed $z = (20\pm1)\,\text{mm}$, with one hit in the pixelated camera from a 10\,min Compton Scanner measurement at a detector temperature of 85\,K. All pulses are time-shifted such that the core pulses are aligned to 50\% of their amplitude.}
	\label{fig:onehit}
\end{figure}

\subsection{Events with one hit in the camera}

Figure~\ref{fig:onehit} shows pulses from events with one hit in the camera from a 10\,min measurement at a typical detector temperature of~85\,K, for which the detector was irradiated at $r~=~33.5\,\text{mm}$ and $\phi~=~46.1^\circ$, i.e.\;close to the $\langle100\rangle$ axis in segment~1. In total, 619 events were reconstructed at $z = (20\pm1)\,\text{mm}$. All pulses were shifted to a zero baseline and the tail of the pulses was corrected for the exponential decay of the charge in the preamplifiers~\citep{Hagemann2020}. In addition, the pulses were normalized by $E_\text{det}$ as measured in the core. The pulses from any given event were collectively time-shifted such that the core pulses reached 50\% of their final amplitude at the same time. Segment~1 is the collecting segment. The other segments show so-called mirror pulses which return to the baseline as soon as the charge carriers are collected at the contacts.

The shapes of the 619 pulses recorded in the core vary significantly. At the same time, the lengths of the pulses recorded in segment~1 show a wide spread. These variations result from misreconstructed events which most probably are multi-site events in the segBEGe detector. Especially the multiple turning points observed in some core pulses are known to be an indication for multi-site events~\citep{GERDA2013,MAJORANA2019_PSA}. As events with only one hit in the camera do not allow for a validation of the reconstructed interaction point, it is expected that some events are assigned to the wrong $z$. The pulses from these events should not enter the averaging to create~superpulses.

\begin{figure}[tp]
	\centering
    \includegraphics[width=\linewidth]{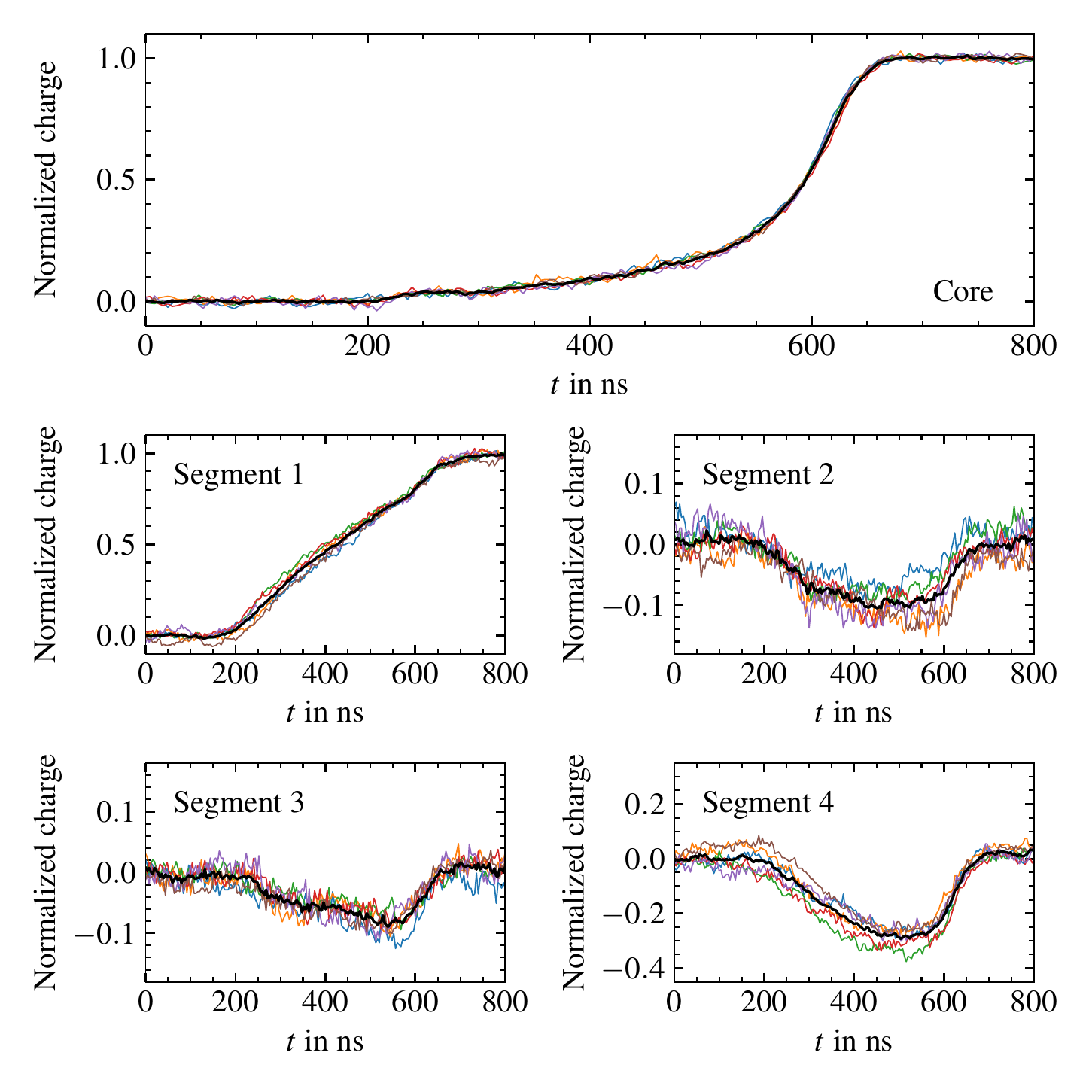}
	\caption{Individual pulses from events assigned to the interaction point at $r = 33.5\,\text{mm}$, \mbox{$\phi = 46.1^\circ$} and a reconstructed \mbox{$z = (20\pm1)\,\text{mm}$}, using only events with two hits in the pixelated camera that passed the Compton cone validation, see Sect.~\ref{subsec:reconstruction}, from a 10\,min Compton Scanner measurement at a detector temperature of 85\,K. All pulses are time-shifted such that the core pulses are aligned to 50\% of their amplitude. Shown in black are the superpulses. 
	}
	\label{fig:twohits}
\end{figure}

\subsection{Events with two hits in the camera}

Events with two hits in the camera, which additionally passed the Compton cone validation presented in Sect.~\ref{subsec:reconstruction}, were collected for all volumes. Figure~\ref{fig:twohits} shows all pulses from validated events with two hits in the camera for the same volume as considered for Fig.~\ref{fig:onehit}. From the same 10\,min measurement, only six pulses remain. In contrast to Fig.~\ref{fig:onehit}, the lengths and shapes of all pulses agree except for statistical fluctuations. This verifies the selection. The solid black lines in Fig.~\ref{fig:twohits} are the superpulses from the six events for the individual contacts.

There are rare cases, in which events are falsely validated. Thus, an additional similarity cut is performed on all events with two hits in the camera. Only events, for which the pulses in the core and the collecting segment agreed within $\chi^2/\text{ndf} \leq 3$ with the respective superpulses formed from all events were kept. Here, $\chi^2/\text{ndf}$ is determined from the 201 samples in the 800\,ns long signal window shown in Fig.~\ref{fig:twohits}:
\begin{linenomath}
\begin{equation}
    \dfrac{\chi^2}{\text{ndf}} = \dfrac{1}{200}\sum\limits_{t = t_0}^{t_0 + 800\,\text{ns}} \dfrac{(q(t) - Q(t))^2}{\sigma^2}
\end{equation}
\end{linenomath}
where $t_0$ is the beginning of the signal window, $q(t)$ and $Q(t)$ are the values of the individual pulse and the superpulse at time $t$, respectively, and $\sigma$ is the root mean square of the individual pulse as determined from the baseline. This similarity cut had no effect on measurements close to the edge but resulted in the rejection of individual events close to the center of the~detector.

\subsection{Final superpulses} \label{subsec:superpulses}

When considering only events with two hits in the camera, the number of pulse shapes is not sufficient to reduce the influence of noise effectively enough by averaging. Although the resulting superpulses are less noisy compared to the individual pulses, the effect of the noise is still too large to observe small features in the pulse shapes, see Fig.~\ref{fig:twohits}. 

The two-hit superpulses were used to identify the misreconstructed events with one hit in the camera.
Many core pulses from events with one hit in the camera are similar to the superpulses obtained from two-hit events. One-hit events were used if individual core and collecting-segment pulses passed a strict similarity cut of $\chi^2/\text{ndf} \leq 1$ with respect to the two-hit event superpulses.

\begin{figure}[b]
	\centering\vspace*{-4pt}
	\includegraphics[width=\linewidth]{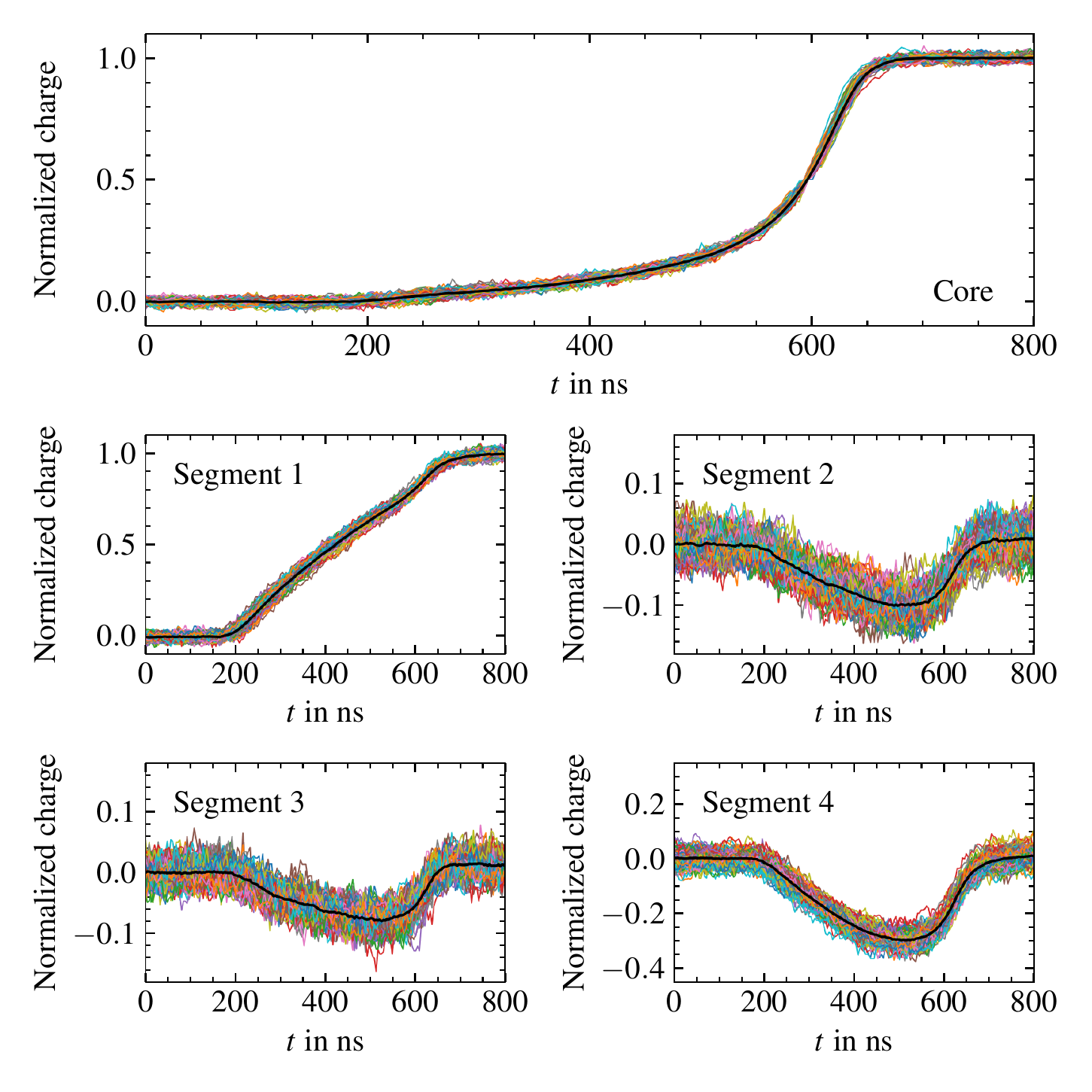}
	\caption{Pulses from events assigned to the interaction point at \mbox{$r = 33.5\,\text{mm}$}, \mbox{$\phi = 46.1^\circ$} and a reconstructed $z = (20\pm1)\,\text{mm}$, with two hits in the camera or with one hit in the camera and selected as described in Sect.~\ref{subsec:superpulses}. All pulses are time-shifted such that the core pulses are aligned to 50\% of their amplitude. Shown in black are the~superpulses.}
	\label{fig:onehit_accepted}
\end{figure}

The selected pulse shapes from events with one hit in the camera, together with the validated two-hit events, are shown in Fig.~\ref{fig:onehit_accepted}. For this measurement, 132 out of the 619 one-hit events survived the similarity cut. By combining the pulses from validated two-hit events and the pulses from similar one-hit events, combined superpulses were obtained for the core and each of the four segments, shown as black lines in~Fig.~\ref{fig:onehit_accepted}. These superpulses are significantly less affected by noise and form a high-quality pulse shape library.

%--------------------------------------------------------

\section{Validation of the Compton reconstruction}

To validate the spatial reconstruction in $z$, pulses at a source position of $\phi = 46.1^\circ$ and $r = 35.5\,\text{mm}$, i.e.\ on the $\langle100\rangle$ axis and 2\,mm away from the mantle surface, were compared to pulses from a measurement where the side of the detector was irradiated with 80.997\,keV gammas from a $^{133}$Ba~source. 

For this measurement, a collimated 1\,MBq $^{133}$Ba source was mounted onto the vertical motor stage of the Compton Scanner frame, see Fig.~\ref{fig:comptonscanner}. Most of the 80.997\,keV gammas interact with germanium via the photoelectric effect with a mean free path of 2.68\,mm~\citep{NIST}. Averaging pulses with an energy of $(81\pm2)\,\text{keV}$ allows to create reference superpulses originating from the first few mm underneath the surface. Note that the distribution of events in the $^{133}$Ba measurement in $r$ is slightly different from that in the Compton Scanner. It decreases exponentially with increasing distance to the surface while the Compton Scanner events are expected to uniformly spread around $r = 35.5\,\text{mm}$ with a FWHM of around 1.4\,mm, see Fig.~\ref{fig:intrinsic_alignment}a.

\begin{figure}[b]
    \centering
    \includegraphics[width=\linewidth]{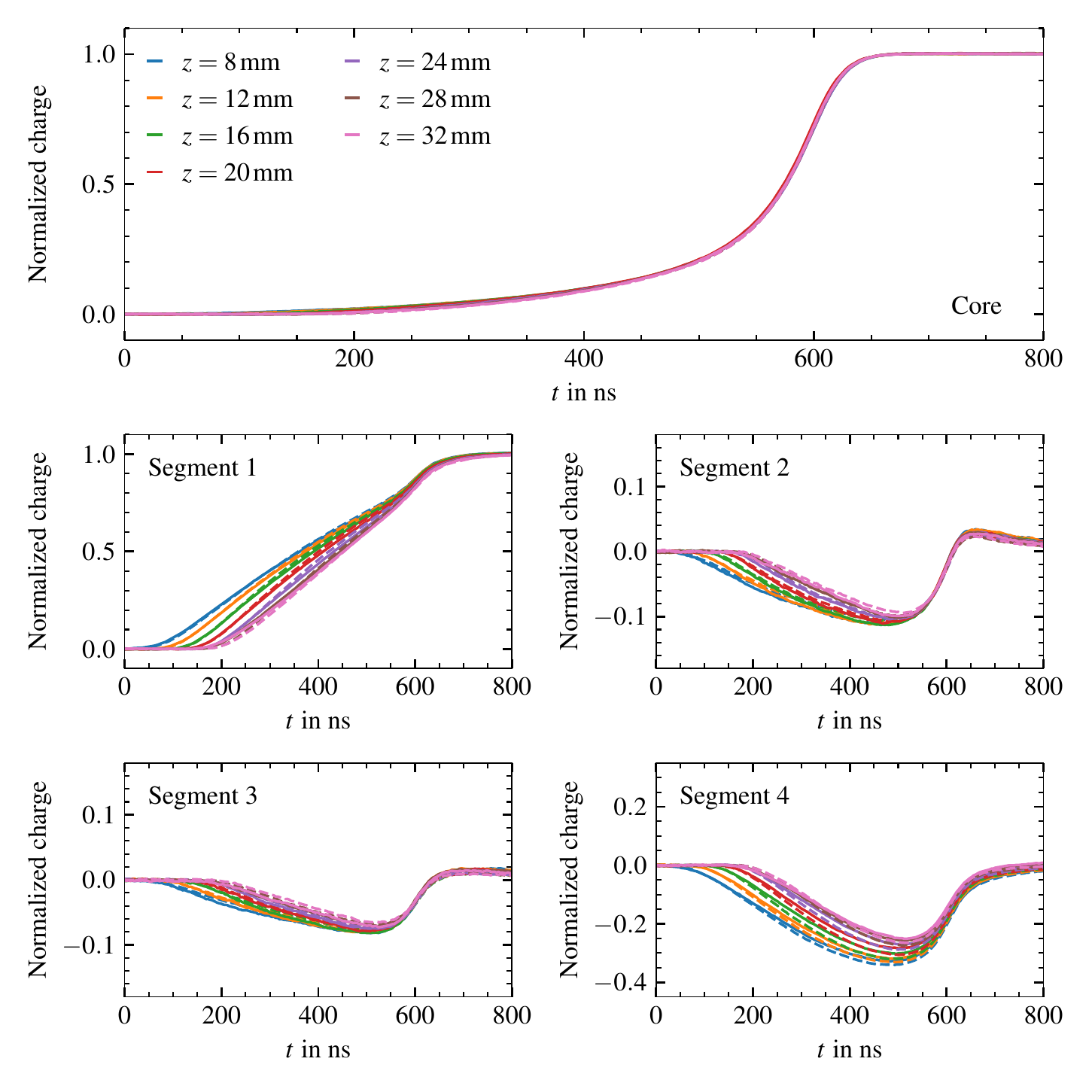}
    \caption{Superpulses of events assigned to interaction points on the $\langle100\rangle$ axis in segment~1 of the segBEGe detector for different $z$ as given in the legend at a detector temperature of~78\,K. The solid lines depict superpulses from a Compton Scanner measurement at $\phi~=~46.1^\circ$ and $r = 35.5\,\text{mm}$. The dashed lines depict $(81\pm2)\,\text{keV}$ superpulses from $^{133}$Ba~measurements at $\phi~=~47.2^\circ$. All pulses are time-shifted such that the core pulses are aligned to 70\% of their amplitude.
    }
    \label{fig:pulseBa}
\end{figure}

Figure~\ref{fig:pulseBa} shows an overlay of Compton superpulses for different $z$ from a 10\,min measurement at a source position of \mbox{$\phi = 46.1^\circ$} and \mbox{$r = 35.5\,\text{mm}$} and \mbox{$(81\pm2)\,\text{keV}$} superpulses from the $^{133}$Ba~measurement where the detector was irradiated from the side at \mbox{$\phi = 47.2^\circ$} at different $z$ in steps of 4\,mm.
In general, the pulses from the different scanning methods are in excellent agreement. The residuals are at the percent level and most probably arise from the $1.1^\circ$ offset in $\phi$, a slightly different distribution of events in $r$, which are used in the superpulse creation, a slightly different cross-talk behavior, as well as an uncertainty of 1\,K on the temperature of the cryostat K2.

\begin{figure}[tb]
    \centering
    \includegraphics{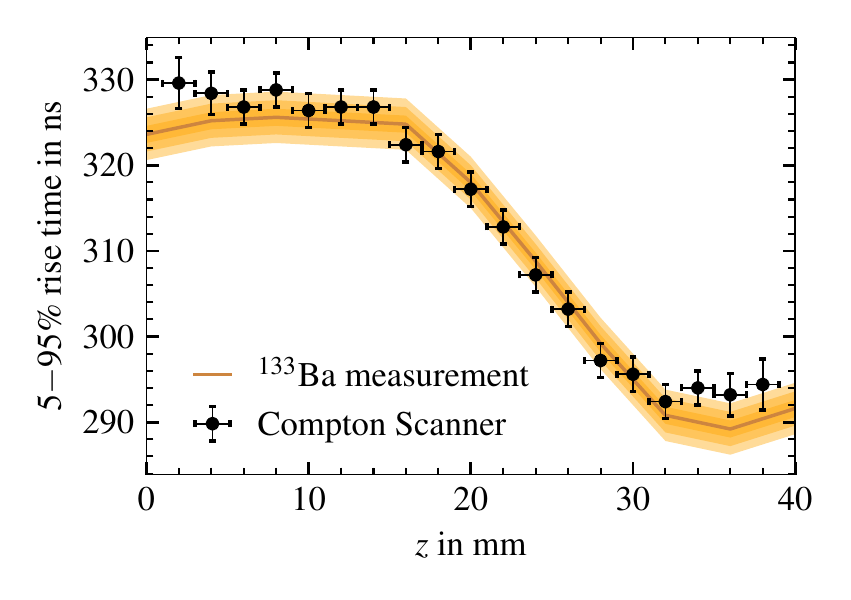}
    \caption{5$-$95\% rise times, $t_r$, as determined from core superpulses for the same source position and detector temperature as the pulses in Fig.~\ref{fig:pulseBa}. Vertical error bars and confidence bands correspond to deviations from $t_r$ when determining the rise times from superpulses of only a subset of pulses. Horizontal error bars correspond the $z$ range used to determine Compton superpulses.
    }
    \label{fig:risetimes}
\end{figure}

The core superpulses in Fig.~\ref{fig:pulseBa} show a small but significant dependence on $z$ at the beginning of the pulse. This difference can be studied in more detail using the 5$-$95\% rise times, $t_r$, i.e.\ the time difference between the pulse reaching 5\% and 95\% of its final amplitude. Due to electronic noise and the sampling time of 4\,ns, determining $t_r$ for individual pulses would lead to uncertainties of the order of 10 to~20\,ns. Thus, $t_r$ is determined from the final superpulses from the Compton Scanner and $^{133}$Ba measurement in Fig.~\ref{fig:pulseBa}, which allows for significantly smaller uncertainties of the order of~2\,ns, see Fig.~\ref{fig:risetimes}. These uncertainties were estimated from the variation of $t_r$ from auxiliary superpulses created from only a subset of the pulses that form the final superpulse.

According to the reference core superpulses from the $^{133}$Ba measurement, $t_r$ only changes significantly in the region of $z$ between 16\,mm and 34\,mm. In this region, $t_r$ of the core superpulses from the Compton Scanner measurement match those from the $^{133}$Ba measurement. Above and below this region, $t_r$ is basically constant and, thus, not sensitive to changes in $z$. This behavior is also well reproduced by the Compton superpulses. At the top and the bottom of the detector, the statistics in Compton Scanner measurements is reduced compared to the middle height of the detector at \mbox{$z = 20\,\text{mm}$}, resulting in slightly noisier pulses. This could explain why the absolute values in those regions $t_r$ are seen to differ slightly from those from the $^{133}$Ba measurement.

In principle, choosing a lower threshold for the rise time than 5\% should increase the sensitivity. However, due to the difference in the spatial distribution of events that enter the superpulse formation, the superpulses from the $^{133}$Ba measurement are expected to be slightly longer in the part of the pulse below the 2\% amplitude. This artifact resulting in a difference in $t_r$ is avoided by setting the lower threshold of the rise time well above 2\%.

In conclusion, the excellent agreement of the pulses in Fig.~\ref{fig:pulseBa} and the core rise times in Fig.~\ref{fig:risetimes} verify the spatial reconstruction in $z$. As an additional cross-check, a second CdZnTe camera, installed at a 45\textdegree{} angle next to the first one, was irradiated from the top with the collimated gamma beam from the $^{137}$Cs source and the events were reconstructed using events with a single hit in each of the cameras. Then, the reconstructed $z_\theta$ were compared to the measured $z$ values from the second camera. Here, the resolution in $\Delta z$ was measured to be $3.2\,\text{mm}$ FWHM. An even better $\Delta z$ resolution is expected for the measurements on detectors installed in the center of the Compton Scanner frame.

\section{Comparison to simulated bulk pulses}

For an additional comparison, pulses originating from the reconstructed volumes were simulated using the \julia{} open-source software package \SSD~\citep{Abt2021}. For the simulation, the impurity density of the germanium detector was set to 90\% of the values given by the manufacturer~\citep{Abt2019}, assuming a linear gradient in $z$, to properly describe the observed depletion voltage. The charges were drifted according to the default electron and hole drift models from \SSD{}~\citep[App.~A]{Abt2021}, which are based on literature values for the electron and hole mobilities along the $\langle100\rangle$ and $\langle111\rangle$~axes at 78\,K~\citep{Bruyneel2006} and theoretical models that determine the anisotropy of the charge drift in between~\citep{Mihailescu2000}. The simulated raw pulses were convolved with the corresponding preamplifier response functions to account for the limited bandwidths of the preamplifiers. In addition, linear and differential cross-talk between the contacts was taken into account~\citep{Hagemann2020}.

Different interaction points in the germanium detector lead to different charge drift paths. Thus, the pulse shapes depend characteristically on the position of the interaction point. In the n-type segBEGe detector, the electrons drift towards the core contact while the holes are collected at the segment contacts. At the center height of the detector, i.e.\;at \mbox{$z=20\,\text{mm}$}, the electric field in the segBEGe detector points primarily inwards in $r$-direction~\citep{Abt2019,Abt2021}. Thus, on the $\langle100\rangle$ axis, the electrons are expected to drift almost horizontally towards the center of the detector along the $[\bar{1}00]$~direction before being pulled upwards towards the core contact along the $[001]$~direction. The holes are expected to drift horizontally~outwards along the $[100]$~direction.

\begin{figure}[t]
    \centering
    \begin{overpic}[width=\linewidth]{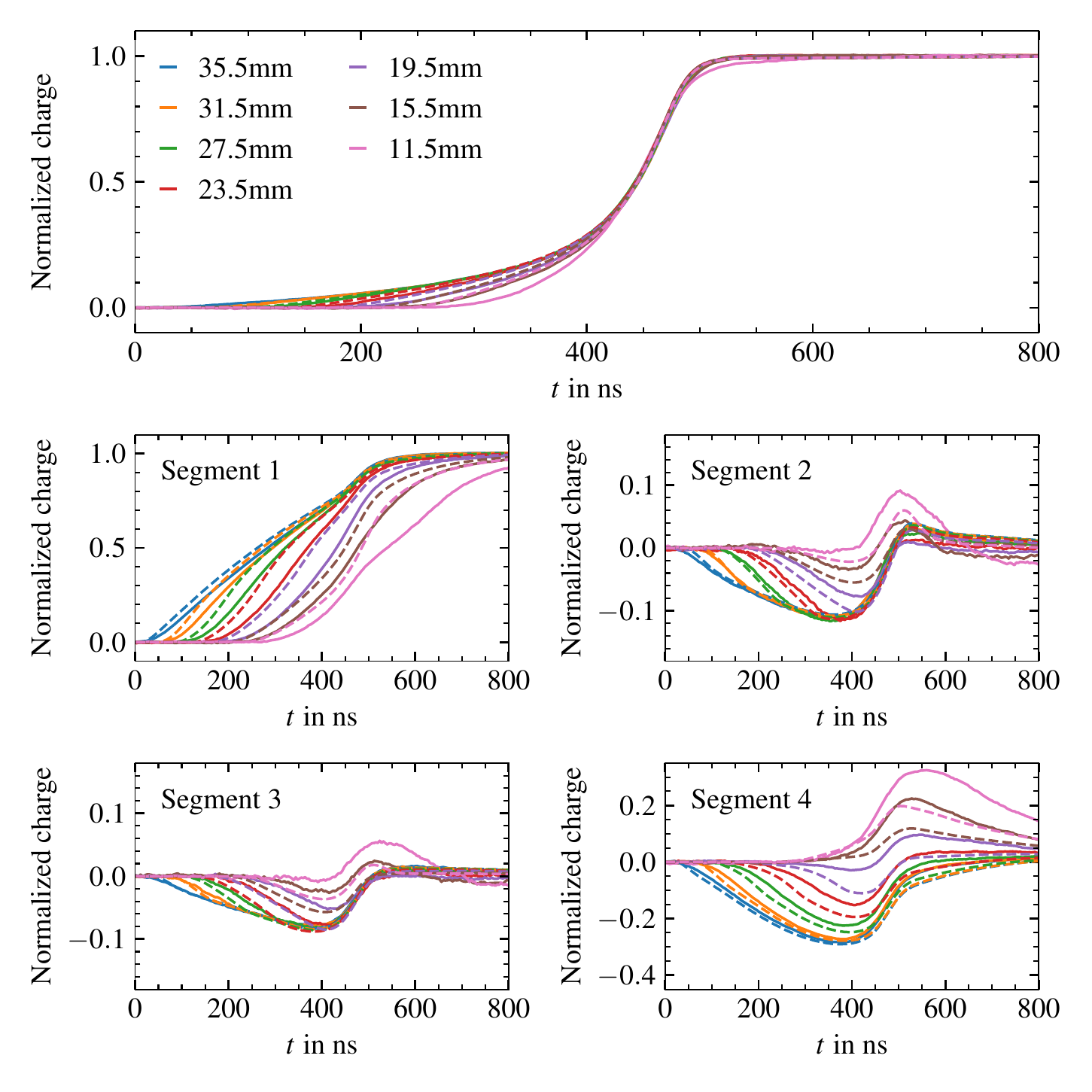}
    \end{overpic}
    \caption{Superpulses of events assigned to interaction points on the $\langle100\rangle$ axis in segment~1 of the segBEGe detector at \mbox{$\phi = 46.1^\circ$}, \mbox{$z = 20\,\text{mm}$} and different $r$ as given in the legend at a detector temperature of~78\,K. The solid lines depict Compton superpulses determined as described in Sect.~\ref{subsec:superpulses}. The dashed lines depict pulses simulated with \SSD~\citep{Abt2021}. For the simulation, the impurity density was set to 90\% of the values given by the manufacturer~\citep{Abt2019}. The preamplifier bandwidths and contributions from linear and differential cross-talk were added to the pulses~\citep{Hagemann2020}. All pulses are time-shifted such that the core pulses are aligned to 70\% of their~\mbox{amplitude.}
    } \label{fig:pulsesim}
\end{figure}

Figure~\ref{fig:pulsesim} shows measured and simulated pulses assigned to interaction points at \mbox{$z=(20\pm1)\,\text{mm}$} on the $\langle100\rangle$ axis in segment~1 for $r$ between~11.5\,mm and~35.5\,mm. Close to the mantle surface, the simulated pulses qualitatively agree with the measured superpulses, especially the one at $r = 35.5\,\text{mm}$, for which the simulation was previously verified using surface events~\cite{Abt2021}. The simulation indicates that the paths of the final upward drift of the electrons are almost equal for all of these events. This results in the last parts of the simulated pulses being equal for all \mbox{$r \geq 11.5\,\text{mm}$}.

The most significant dependence of the core pulse shapes on $r$ is seen at the beginning of the rise of the pulses. As $r$ decreases, the lengths of the pulses decrease. This is due to the electron drift paths getting shorter and the electrons being collected earlier. Close to the mantle surface, the change in the shape of the first parts of the pulses of the collecting and non-collecting segments is well described by the simulation. The observation that the characteristic rise at the end of the core pulses is identical for all \mbox{$r \geq 15.5\,\text{mm}$} implies that the holes are already collected at the segment contacts before the electrons reach the vicinity of the core~contact. However, at \mbox{$r = 11.5\,\text{mm}$}, the final part of the pulse deviates from the characteristic shape observed for larger $r$, implying that the electrons are collected at the core contact before the holes reach the segment contacts. This is not predicted by the simulation. For \mbox{$r \leq 7.5\,\text{mm}$}, the holes drift mostly downwards and are collected on the closed bottom end-plate of segment~4 instead of on segment~1.

With decreasing $r$, the predictions for the pulse lengths and mirror pulse amplitudes disagree more from the data, implying that the charge drift in the bulk of the detector is not well modeled by the simulation. Especially for events at \mbox{$r \leq 19.5\,\text{mm}$}, where the outwards drift of the holes has a larger contribution, the amplitudes of the mirror pulses are smaller than predicted by the simulation. This shows the power of the Compton Scanner setup presented in this paper to test the models used in the simulation, e.g.~the assumed model for the impurity density profile of the germanium crystal and the charge drift models for electrons and~holes.

%--------------------------------------------------------

\section{Summary and outlook} \label{sec:summary}

A novel Compton Scanner setup to investigate bulk events in germanium detectors has been presented in this paper. It is based on detecting Compton scattered gammas using an energy- and position-sensitive camera, which allows for detecting a wider range of Compton angles in comparison to Compton Scanner setups using slit collimators.

The Compton Scanner was commissioned with an n-type segmented point-contact germanium detector, for which pulse shape libraries at different temperatures were collected. Monte Carlo simulations indicate that pulses in the bulk of the detector can be successfully reconstructed with a FWHM resolution of less than $\pm1\,\text{mm}$ along the beam direction. The spatial reconstruction was validated by a comparison with measured surface pulses from a $^{133}$Ba measurement. In addition, pulses measured with the Compton Scanner were compared to bulk pulses simulated using \SSD{} for drifts along the $\langle100\rangle$ crystallographic axis at the middle height of the detector. For such drifts, the simulation is most reliable. The simulated pulses describe the trends of the measured pulses close to the surface quite well. Deviations for pulses resulting from events in the bulk of the detector indicate that simulations need to be tested and tuned with the data from the Compton Scanner. A systematic analysis is planned for all axes and for different temperatures.

In order to further decrease the measurement times, the Compton Scanner will be upgraded by installing a second pixelated camera at a 45\textdegree{} angle next to the first~one. This will increase the statistics by almost a factor of two. Future Compton Scanner measurement campaigns are planned with a p-type segmented point-contact detector~\citep{Hagemann2020}, an inverted coaxial point-contact detector~\citep{GERDA2021} and a segmented true-coaxial detector~\citep{Abt2007B}. Pulse shape libraries from different detectors will also be used to characterize pulse shape discrimination algorithms for rare-event searches using germanium detectors.

%--------------------------------------------------------

\section*{Acknowledgments}

The authors would like to thank Dr.\ Hao Yang from H3D,~Inc.\ for his valuable feedback and support on operational questions regarding the pixelated CdZnTe camera and to thank Nuno Barros for very fruitful discussions during the early conception of the experimental approach presented here.

%--------------------------------------------------------

\appendix
\renewcommand{\thefigure}{A\arabic{figure}}
\renewcommand{\theHfigure}{A\arabic{figure}}
\setcounter{figure}{0}

\section{Spatial alignment}\label{app:alignment}

\begin{figure}[tb]
    \centering
    \includegraphics[width=\linewidth]{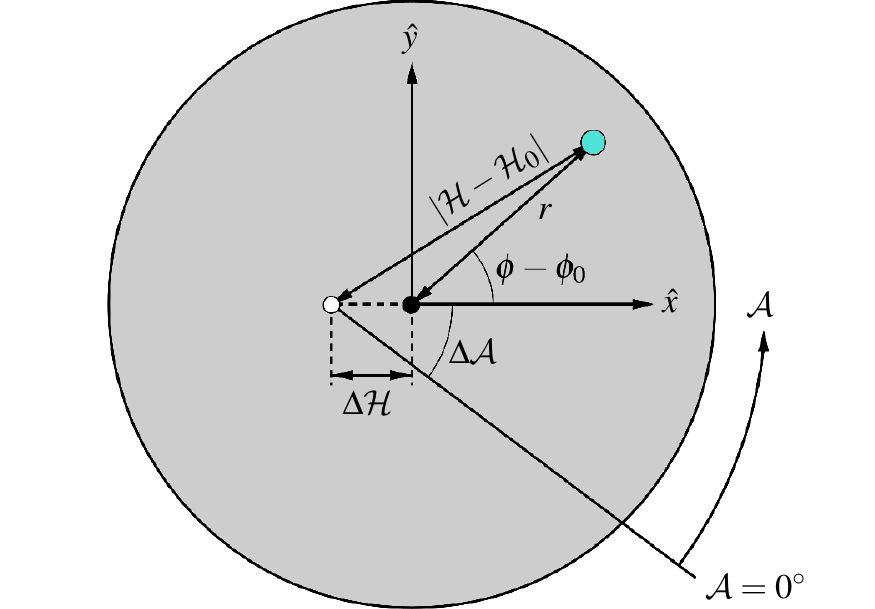}\\[10pt]
    \caption{Schematic of the coordinate systems used to describe positions of the vertical beam in the horizontal plane. The center of the Compton Scanner frame (white point) and the center of the detector (black point) have an offset of $\Delta\mathcal{H}$, which is reached at \mbox{$\mathcal{A} = \Delta\mathcal{A}$} or~\mbox{$\phi = \phi_0$.}}
    \label{fig:appendixA}
\end{figure}

Figure~\ref{fig:appendixA} depicts the different coordinate systems used to describe the position of the vertical beam in the horizontal plane: Cartesian coordinates, $\hat{x}$ and $\hat{y}$, detector coordinates, $r$ and $\phi$, and motor coordinates, $\mathcal{H}$ and $\mathcal{A}$.

The Cartesian coordinate system is defined with the origin at the center of the detector (black point in Fig.~\ref{fig:appendixA}) and the $\hat{x}$~axis pointing parallel to the line that connects the center of the detector and the center of the Compton Scanner frame (black and white points in Fig.~\ref{fig:appendixA}). The $\hat{x}$~axis does not necessarily need to be aligned with $\phi = 0^\circ$, see Sect.~\ref{subsec:segBEGe}, where $\phi = 0^\circ$ is defined as the segment boundary between segments 1 and~4. In the detector coordinate system, the $\hat{x}$~axis is defined to be aligned with $\phi = \phi_0$. In the motor coordinate system, the $\hat{x}$~axis is aligned with $A = \Delta\mathcal{A}$. This yields the following relations between the different coordinates:
\begin{linenomath}
\begin{gather}
\hat{x} = r \cos(\phi - \phi_0) = \vert\mathcal{H} - \mathcal{H}_0\vert \cos(\mathcal{A} - \Delta \mathcal{A}) - \Delta\mathcal{H} \label{eq:x}\\
\hat{y} = r \sin(\phi - \phi_0) = \vert\mathcal{H} - \mathcal{H}_0\vert \sin(\mathcal{A} - \Delta \mathcal{A})~. \label{eq:y}
\end{gather} 
\end{linenomath}

Combining Eqs.~\eqref{eq:x} and \eqref{eq:y} results in expressions to convert motor coordinates to detector coordinates:
\begin{linenomath}
\begin{gather}
r^2 = \vert\mathcal{H} - \mathcal{H}_0\vert^2 + (\Delta\mathcal{H})^2 - 2 \Delta \mathcal{H} \vert\mathcal{H} - \mathcal{H}_0\vert \cos(A - \Delta\mathcal{A}) \label{eq:r}\\
\phi - \phi_0 = \arcsin\left(\dfrac{\vert\mathcal{H} - \mathcal{H}_0\vert \sin(\mathcal{A} - \Delta \mathcal{A})}{r}\right)~, \label{eq:phi}
\end{gather}
\end{linenomath}
with $\phi_0 = \Delta \mathcal{A} - \mathcal{A}_0$, where $\mathcal{A}_0$ is the offset between $\phi = 0^\circ$ and $\mathcal{A} = 0^\circ$, see Sect.~\ref{sec:alignment}. For $\Delta\mathcal{H} \rightarrow 0$, Eq.\;\eqref{eq:r} becomes \mbox{$r = \vert\mathcal{H} - \mathcal{H}_0\vert$} and Eq.\;\eqref{eq:phi} becomes \mbox{$\phi - \phi_0 = \mathcal{A} - \Delta\mathcal{A}$}, \linebreak i.e.\;\mbox{$\phi = \mathcal{A} - \mathcal{A}_0$}, which is what is expected for perfectly aligned centers.

To determine the misalignment quantities, $\Delta\mathcal{H}$ and $\Delta\mathcal{A}$, and the offsets, $\mathcal{H}_0$ and $\mathcal{A}_0$, from data, the position of the detector edge was determined in horizontal motor coordinates, $\mathcal{H}_e$, for different rotational motor positions, $\mathcal{A}$, see Sect.~\ref{sec:alignment}.
For a cylindrical detector, $r(\mathcal{H}_e, \mathcal{A}) = R$, where $R$ is the constant radius of the detector. Plugging this into Eq.\;\eqref{eq:r} yields the expression for $\mathcal{H}_e < \mathcal{H}_0$ shown in Eq.\;\eqref{eq:HeR}:
\begin{linenomath}
\begin{gather}
R^2 = \vert\mathcal{H}_e - \mathcal{H}_0\vert^2 + (\Delta\mathcal{H})^2 - 2 \Delta \mathcal{H} \vert\mathcal{H}_e - \mathcal{H}_0\vert \cos(A - \Delta\mathcal{A}) \notag \\
= (\vert\mathcal{H}_e - \mathcal{H}_0\vert - \Delta\mathcal{H}\cos(\mathcal{A} - \Delta\mathcal{A}))^2 + (\Delta\mathcal{H})^2 \sin^2(\mathcal{A} - \Delta\mathcal{A}) \notag \\[10pt]
%\vert\mathcal{H}_e - \mathcal{H}_0\vert - \Delta\mathcal{H}\cos(\mathcal{A} - \Delta\mathcal{A}) = \pm\sqrt{R^2 - (\Delta	\mathcal{H})^2 \sin^2(\mathcal{A} - \Delta\mathcal{A})} \notag \\
\begin{split}
\overset{\scriptstyle\mathcal{H}_e < \mathcal{H}_0}{\Longrightarrow} \mathcal{H}_e(\mathcal{A}) = \mathcal{H}_0 &- \Delta \mathcal{H} \cos(\mathcal{A} - \Delta\mathcal{A}) \\ &\pm \sqrt{R^2 - (\Delta \mathcal{H})^2 \sin^2(\mathcal{A} - \Delta\mathcal{A})}.
\end{split} \label{eq:HeR_new}
\end{gather}
\end{linenomath}
For non-cylindrical detectors, the radius $R$ is not constant and \mbox{$r(\mathcal{H}_e, \mathcal{A}) = R(\phi(\mathcal{H}_e, \mathcal{A})) = R(\mathcal{H}_e, \mathcal{A})$} is a function of $\mathcal{H}_e$ and $\mathcal{A}$. Then, Eq.\;\eqref{eq:HeR_new} is an implicit function in $\mathcal{H}_e$.

%--------------------------------------------------------

\bibliographystyle{spphys.bst}
\bibliography{ComptonScanner.bib}

\end{document}